\documentclass{article}

\usepackage[letterpaper,top=2cm,bottom=2cm,left=3cm,right=3cm,marginparwidth=1.75cm]{geometry}

\usepackage{amsmath}
\usepackage{amssymb}
\usepackage{graphicx}
\usepackage[dvipsnames]{xcolor}
\usepackage{subcaption}
\usepackage{tabularx}
\usepackage{dsfont}
\usepackage{bm}
\usepackage{authblk}
\usepackage[colorlinks=true, allcolors=blue]{hyperref}

\title{Escape dynamics and implicit bias of one-pass SGD in overparameterized quadratic networks}

\author[1,2]{Dario Bocchi\thanks{Corresponding author: \texttt{dario.bocchi@uniroma1.it}}}
\author[3]{Theotime Regimbeau}
\author[3,4]{Carlo Lucibello}
\author[3,4]{Luca Saglietti}
\author[1,5]{Chiara Cammarota}

\affil[1]{Physics Department, Sapienza University of Rome, Piazzale Aldo Moro 5, 00185, Rome, Italy}
\affil[2]{Institute of Nanotechnology, CNR-NANOTEC, Piazzale Aldo Moro 5, 00185 Roma RM, Italy}
\affil[3]{Department of Computing Sciences, Bocconi University, 20136 Milano, Italy}
\affil[4]{Institute for Data Science and Analytics, Bocconi University, 20136 Milano, Italy}
\affil[5]{National Institute for Nuclear Physics, INFN-Roma1, Piazzale Aldo Moro 5, Rome, 00185, Italy}

\date{} % leave empty for arXiv

\begin{document}
\maketitle

\begin{abstract}
We analyze the one-pass stochastic gradient descent dynamics of a two-layer neural network with quadratic activations in a teacher–student framework. In the high-dimensional regime, where the input dimension \(N\) and the number of samples \(M\) diverge at fixed ratio \(\alpha = M/N\), and for finite hidden widths \((p,p^*)\) of the student and teacher, respectively, we study the low-dimensional ordinary differential equations that govern the evolution of the student–teacher and student–student overlap matrices. We show that overparameterization (\(p>p^*\)) only modestly accelerates escape from a plateau of poor generalization by modifying the prefactor of the exponential decay of the loss. We then examine how unconstrained weight norms introduce a continuous rotational symmetry that results in a nontrivial manifold of zero‐loss solutions for \(p>1\). From this manifold the dynamics consistently selects the closest solution to the random initialization, as enforced by a conserved quantity in the ODEs governing the evolution of the overlaps. Finally, a Hessian analysis of the population-loss landscape confirms that the plateau and the solution manifold correspond to saddles with at least one negative eigenvalue and to marginal minima in the population-loss geometry, respectively.
\end{abstract}

\section{Introduction}
%\href{https://www.overleaf.com/learn}{help library}
%\href{https://www.overleaf.com/user/subscription/plans}{choose your plan}
%Figure~\ref{fig:frog} 
%\begin{figure}
%\centering
%\includegraphicxs[width=0.25\linewidth]{frog.jpg}
%\caption{\label{fig:frog}This frog was uploaded via the file-tree menu.}
%\end{figure}
%Table~\ref{tab:widgets}
%\begin{table}
%\centering
%\begin{tabular}{l|r}
%Item & Quantity \\\hline
%Widgets & 42 \\
%Gadgets & 13
%\end{tabular}
%\caption{\label{tab:widgets}An example table.}
%\end{table}
%\href{https://www.overleaf.com/learn/latex/page_size_and_margins}{page size and margins}.
%\href{https://www.overleaf.com/learn/latex/International_language_support}{international language support}.

We focus on the instance of a simple machine learning problem involving the training of a two-layer neural network with quadratic activation function and quadratic loss. The training data is generated by a "teacher" function \cite{engel2001statistical} with a similar architecture but possibly a different number of hidden neurons. In this setup, we systematically vary the number of hidden units $p$ in the student and $p^*$ in the teacher to study two relevant aspects of many modern machine learning settings:
\begin{enumerate}
\item overparameterization: By fixing 
$p^*$ and varying $p$, we examine how increasing the student model capacity affects learning.
\item Symmetry and invariance: When the student has multiple hidden units ($p>1$), we investigate the implications of learning with unconstrained norms when the student's function is invariant under continuous transformations. %—i.e., possessing internal symmetries.
\end{enumerate}
The learning setup we consider has already been extensively studied in a number of relevant limiting cases. For $p=p^*=1$ the problem reduces to the well-known and real-valued version of the inference problem of Phase Retrieval, with applications spanning signal processing, imaging, and quantum mechanics \cite{dong2023phase}. 
%X-ray crystallography [Mil90,Har93], microscopy [MISE08], astronomy [FD87], optics [Wal63], acoustics [BCE06], interferometry [DJ17], and quantum mechanics [Cor06].
Phase Retrieval is famously known as a hard problem of reconstruction of a hidden signal, a real or complex vector in $N$ dimensions, represented by the teacher in our description. %Learning in this case can be recast in terms of a non-convex optimization problem which is hard for descent-based algorithms
A plethora of algorithms have been studied and developed to reconstruct the signal in polynomial time~\cite{candes2015phase, netrapalli2013phase, waldspurger2015phase, chen2017solving, zhang2017nonconvex, wang2017solving, wang2017solving2, zhang2018two}, with the goal of reducing as much as possible the number of examples $M$ needed for reconstruction. In particular, no algorithm can guess better than a dummy predictor for $M< 0.5 N$ \cite{barbier19}. Still, in the scarce number of samples regime, suitably defined empirical risk functions are typically very rough so that standard gradient-based algorithms are doomed to fail. %On the other hand a full reconstruction (strong recovery) can be achieved by the approximate message passing algorithm for $M>1.13 N$\cite{} [let's add the new result from improved AMP?].
On the other hand, stochastic gradient descent (SGD) algorithms could do better. In particular, for small learning rate, one-pass SGD can be seen \cite{robbins1951stochastic} as a discretisation of gradient flow on the population risk, which is expected to be devoid of spurious minima otherwise generated by undersampling.   
In both contexts, it is interesting to explore the effect of overparameterization, by increasing $p$ while keeping $p^*=1$, and characterize the easing of the learning effort, as generally expected in realistic applications \cite{novak2018sensitivity, belkin2019reconciling, simon2023more}. 
Gradient flow in the large $p$ limit has been explored in Ref. \cite{sarao2020optimization}: spurious minima in the empirical risk disappear, making the learning task very simple.
The case of one-pass SGD with $p^*=1$ and finite $p>1$ has also been studied \cite{arnaboldi2023escaping}, showing that during the learning dynamics, it is still hard to escape from the initial uninformed condition due to an essentially flat landscape. overparameterization is found to be beneficial in decreasing the escaping time from the uninformative region, but it remains of little help as it only affects the learning time by a prefactor.
More generally, the target function with $p^*=1$ belongs to the class of single-index or generalised linear models, a simple but rich setting that has allowed several recent theoretical progress \cite{tan2023online, arous2021online, bietti2022learning, damian2022neural, berthier2024learning, damian2024computational, montanari2025dynamical}, including the characterization of notable mean field learning regimes \cite{chizat2018global, rotskoff2018neural, mei2018mean, mei2019mean, sirignano2020mean}. 

Also the $p^* >1$ case has already been considered in the literature. Some works have studied approximate message passing algorithms in the Bayes-optimal setting for $p^* >1$ \cite{aubin2018committee}.
Very recently, the case $p^*=p$ and proportional to the input dimension has been considered, with number of samples both proportional~\cite{cui2023bayes} and quadratic~\cite{maillard2024bayes} in the input dimensions.
In broader terms, when \(p^*>1\) and varying \(p\), the setting becomes that of a multi-index model for which ordinary differential equations (ODEs) for online learning can be generalized, and their convergence can be proven for any activation function \cite{collins2024hitting}. 
%The time complexity of a two-time procedure for learning both weights and activations themselves has also been studied \cite{bietti2023learning}. 

In this work, we study online learning of one-hidden layer neural networks with quadratic activations,  in the case $p^*>1$ and generic $p$. We extend the analysis of \cite{arnaboldi2023escaping} to determine how overparameterization affects both the convergence time and the geometry of the loss landscape in a setting where learning can end in a continuous family of equivalent solutions.
%about the change in the structure of the loss landscape and the decrease of the reconstruction time with overparameterization, and we explore the fate of learning when a continuous family of equivalent solutions exists.
In particular, the golf course picture emerging for $p^*=1$ \cite{arnaboldi2023escaping} with a single deep minimum at the solution, is replaced for $p>1$, $p \geq p^*$ and unconstrained norms by a lake landscape: the zero-loss set becomes a continuous manifold of degenerate solutions generated by the underlying rotational symmetry of the model. %, with multiple degenerate solutions, connected by the continuous transformation generated by a class of rectangular orthogonal matrices. 
In this respect, the case $p>1$ captures a feature also observed in more realistic machine-learning settings \cite{draxler2018essentially, baity2018comparing, baldassi2020shaping}, where low-loss regions can be extended and flat, %shares a stronger similarity with more realistic machine learning settings \cite{draxler2018essentially, baity2018comparing, baldassi2020shaping} displaying a flat low-loss region in the loss landscape, 
either as a direct consequence of weight-space symmetries %https://openreview.net/attachment?id=rkxmPgrKwB&name=original_pdf 
or emerging in the overparameterization regime. 
Moreover, this analytically tractable framework allows us to characterize how the initialization selects one learned solution among the many equivalent minima, linking our results to the literature on the implicit bias of gradient-based optimization \cite{gunasekar2017implicit,gunasekar2018characterizing,soudry2018implicit,chizat2020implicit}, as we discuss further below.

\section{The model}

% \textit{In this chapter, we present the model and highlight the main differences with previous works.\\}

% \cc{TO DO if needed, move general concepts in the intro, here leave only the technical description of the model} We extend the analysis of the teacher-student setup for two-layer networks with quadratic activation functions. Our primary contribution is the study of a more general case where the number of perceptrons in the hidden layer of both the teacher and student networks is finite, but greater than one.  This approach allows us to explore the effects of over-parameterization in the student network, as previously studied in other works \cite{du2018power} \cite{martin2024impact}. Moreover, by allowing the hidden layer of the teacher network to have multiple directions (i.e., multiple relevant features), we generalize beyond the classical phase retrieval setup, for which various algorithms have been developed over the years \cite{fienup1982phase}. This modification makes the model more representative of practical machine learning tasks, where the presence of multiple influential features is common and critical to the learning process.\\
In the teacher-student framework, the goal is to optimize a student network $\hat{y}(\vec{x}) : \mathbb{R}^N \rightarrow \mathbb{R}$ such that it approximates the behavior of an unknown teacher network $y(\vec{x}): \mathbb{R}^N \rightarrow \mathbb{R}$. The training process relies on the teacher's outputs evaluated on $M = \alpha N$ input vectors $\{\vec{x}^\mu\}_{\mu =1}^M$, for us randomly and independently drawn from the $N$-dimensional standard Gaussian. For a one-hidden-layer neural network with quadratic activation, the teacher network's output in correspondence of a given input $\vec{x}$ is
\begin{equation}
    y(\Vec{x}) = \frac{1}{p^{*}}\sum_{l = 1}^{p^{*}}\left( \frac{\Vec{w}_{l}^{*} \cdot \Vec{x}}{\sqrt{N}} \right)^2 = \frac{1}{p^{*}}\sum_{l = 1}^{p^{*}}\left(u_{l} \right)^2,
\end{equation}
where $p^*$ denotes the hidden layer width of the teacher network and where the pre-activation output of the $l$-th teacher perceptron $\vec{w}^{*}_l \in \mathbb{R}^N$ is $u_{l} = \Vec{w}_{l}^{*} \cdot \Vec{x}/{\sqrt{N}}$. In the following, we will assume that the teacher's perceptrons are orthogonal to one another, and we normalize their quadratic norms to $N$ ($
    \Vec{w}_{l}^{*} \cdot \Vec{w}_{l'}^{*} = N\delta_{l, l'}$).
Similarly to the teacher network, the output of the student network for a given $\vec{x}$ is
\begin{equation}
    \hat{y}(\Vec{x}) = \frac{1}{p}\sum_{k = 1}^{p}\left( \frac{\Vec{w}_{k} \cdot \Vec{x}}{\sqrt{N}} \right)^2 = \frac{1}{p}\sum_{k = 1}^{p}\left(\lambda_{k} \right)^2,
\label{eq:studentOutput}
\end{equation}
where $p$ denotes the hidden layer width of the student network and where the pre-activation output of the $k$-th student perceptron $\vec{w}_k \in \mathbb{R}^N$ is defined as $\lambda_{k} = \Vec{w}_{k} \cdot \Vec{x}/{\sqrt{N}}$. 

The optimal configuration of the student weights $w=\{\vec{w}_{k}\}_{k=1}^{p}$ is usually obtained by minimizing the empirical risk function $\mathcal{L}$
, defined as the sum of the squared differences between the outputs of the student and teacher networks over all the training data points:
\begin{equation}
    \mathcal{L}(w) = \sum_{\mu=1}^{M}\ell(\Vec{x}^{\mu}) = \sum_{\mu=1}^{M} \left( y(\Vec{x}^{\mu}) -  \hat{y}(\Vec{x}^{\mu}) \right)^2 = \sum_{\mu=1}^{M}\left(\frac{1}{p^{*}}\sum_{l = 1}^{p^{*}}\left(u^{\mu}_{l} \right)^2 -  \frac{1}{p}\sum_{k = 1}^{p}\left(\lambda^{\mu}_{k} \right)^2\right)^2,
    \label{eqn:loss}
\end{equation}
with $u_{l}^\mu = \Vec{w}_{l}^{*} \cdot \Vec{x}^\mu/{\sqrt{N}}$ and $\lambda^\mu_{k} = \Vec{w}_{k} \cdot \Vec{x}^\mu/{\sqrt{N}}$. 
The performance of the network is then evaluated using the population risk $\mathcal{R}$, which measures the expected loss over the entire population of input data: $\mathcal{R}(w)= \mathbb{E} \left[\ell(\Vec{x})\right]$, where the expectation is taken over the input distribution. 

\section{Online learning}

In this work, the student network's weights are optimized via one-pass stochastic gradient descent (SGD), also called \textit{online learning}, where each training sample $\vec{x}^{\mu}$ is processed exactly once. This framework differs from classical batch methods by establishing a direct correspondence between weight update and processed sample, both indexed by $\mu$.  For each presented example $\vec{x}^{\mu}$, the weights update is
\begin{equation}
    \Vec{w}_{k}^{\mu + 1} - \Vec{w}_{k}^{\mu} = -\eta \Vec{\nabla}_{\Vec{w}_{k}}\ell(\Vec{x}^{\mu}) = \frac{1}{\sqrt{N}} F_{k}^{\mu} \Vec{x}^{\mu}, 
    \label{eqn:SGD}
\end{equation}
where $\eta$ is the learning rate and $F_{k}^{\mu}$ is called the learning amplitude and defined as
\begin{equation}
    F_{k}^{\mu} = \frac{4 \eta}{p} \left(\frac{1}{p^{*}}\sum_{l=1}^{p^{*}}(u^{\mu}_{l})^2 - \frac{1}{p}\sum_{r=1}^{p}(\lambda^{\mu}_{r})^2\right)\lambda_{k}^{\mu}.
    \label{eq:update_amplitude}
\end{equation}

We analyze the high-dimensional limit where the input size $N$ and the total number of examples $M$ both go to infinity, while keeping $\alpha = M/N$ finite and with finite hidden layer sizes ($p, p^{*} = \mathcal{O}(1)$). The dynamics is governed by two distinct types of order parameters: the teacher-student overlap matrix $\boldsymbol{\rho} \in \mathbb{R}^{p \times p^{*}}$ with elements $\rho_{kl} = \Vec{w}_{k} \cdot \Vec{w}_{l}^{*}/N$, and the student-student overlap matrix $\mathbf{Q} \in \mathbb{R}^{p \times p}$ with elements $Q_{kk'} = \Vec{w}_{k} \cdot \Vec{w}_{k'}/N$, where $k,k' =1, \ldots, p$ and $l =1, \ldots, p^*$. \\

Following the seminal work of Ref. \cite{saad1995line} (later made rigorous in Ref. \cite{goldt2019dynamics}), one can show that the order parameters evolve according to deterministic ordinary differential equations (ODEs):
\begin{equation}
    \frac{d\rho_{kl}}{d\alpha} = \langle F_{k} u_{l} \rangle, \quad
    \frac{dQ_{kk'}}{d\alpha} = \langle F_{k} \lambda_{k'} \rangle + \langle F_{k'} \lambda_{k} \rangle + \langle F_{k} F_{k'} \rangle,
    \label{eq:ode_system}
\end{equation}

where $\alpha$ plays the role of time and the averages $\langle \cdot \rangle$ are computed with respect to a jointly Gaussian distribution $P(\{u_{l}\}_l, \{\lambda_{k}\}_k)$ (see Appendix \ref{sec:complete_equations} for the complete equations). This distribution has zero mean and covariance structure given by:

\begin{equation}
\mathbf{\Sigma} = \begin{bmatrix}
\mathbf{I}_{p^{*}} & \boldsymbol{\rho}^T \\
\boldsymbol{\rho} & \mathbf{Q} \\
\end{bmatrix},
\end{equation}

with $\mathbf{I}_{p^{*}}$ denoting the $p^{*}\times p^{*}$ identity matrix, $\boldsymbol{\rho} \in \mathbb{R}^{p\times p^{*}}$ the teacher-student overlap matrix, and $\mathbf{Q} \in \mathbb{R}^{p\times p}$ the student-student overlap matrix defined above.

% \begin{equation}
% \mathbf{\Sigma} =  \begin{bmatrix}
%     1      & 0      & \dots  & 0       & \rho_{11} & \rho_{21} & \dots & \rho_{p1} \\
%     0      & 1      & \dots  & 0       & \rho_{12} & \rho_{22} & \dots & \rho_{p2} \\
%     \vdots & \vdots & \ddots & \vdots  & \vdots    & \vdots    & \ddots&  \vdots   \\
%     0      & 0      & \dots  & 1       &  \rho_{1p^{*}} & \rho_{2p^{*}} & \dots & \rho_{pp^{*}}\\

%     \rho_{11} & \rho_{12} & \dots & \rho_{1p^{*}} & Q_{11} & Q_{12} & \dots & Q_{1p}\\
%     \rho_{21} & \rho_{22} & \dots & \rho_{2p^{*}} & Q_{12} & Q_{22} & \dots & Q_{2p}\\
%     \vdots    & \vdots    & \ddots& \vdots    & \vdots & \vdots & \ddots&\vdots \\
%     \rho_{p1} & \rho_{p2} & \dots & \rho_{pp^{*}} & Q_{1p} & Q_{2p} & \dots & Q_{pp}\\
% \end{bmatrix} = \begin{bmatrix}
% \mathbf{I} & \boldsymbol{\rho}^T \\
% \boldsymbol{\rho} & \mathbf{Q} \\
% \end{bmatrix}.
% \end{equation}

\paragraph{Initialization of the student network} In online learning, a common practice is to initialize the student model in a \textit{tabula rasa} state with $\vec{w}_k(0) = \vec{0}$ for all $k$. However, for quadratic activation functions, this initialization leads to a trivial fixed point since the learning amplitude in Equation~\eqref{eq:update_amplitude} is proportional to the student's pre-activation outputs. To enable non-trivial learning dynamics, the weights must instead be initialized with non-zero values.

A second trivial fixed point occurs when the teacher and student networks are initially orthogonal ($\rho_{kl}(0) = 0 \; \forall k, l$). In this case, the symmetry of the system causes the teacher-student overlap evolutions $d\rho_{kl}/d\alpha$ in Equation~\eqref{eq:ode_system} to vanish identically $ \forall k, l$. In finite-$N$ implementations with $\|\vec{w}^{*}_l(0)\|^2 = \|\vec{w}_k(0)\|^2 = {N}$, random initialization typically yields small but non-zero initial overlaps of order $\rho_{kl} \sim \mathcal{O}(1/\sqrt{N})$ between student and teacher weights, which breaks this symmetry and allows for meaningful learning.

To capture this essential feature of finite-dimensional systems while working in the $N\to\infty$ limit, we initialize the student nodes to be mutually orthogonal with quadratic norm equal to $N$, {\it i.e.} $\vec{w}_k (0)~\cdot~\vec{w}_{k'}(0) = N\delta_{kk'}$, while introducing small random overlaps with the teacher nodes. Specifically, we will consider random initial conditions where each $\rho_{kl}$ is independently sampled from a Gaussian distribution of zero mean and $\epsilon^2$ variance:

\begin{equation}
    \begin{cases}
      \rho_{kl}(0) \overset{\text{i.i.d.}}{\sim} \mathcal{N}(0, \epsilon^2), & k=1, \ldots, p, \; l=1, \ldots, p^* \\
      Q_{kk'}(0) = \delta_{kk'}, & \forall k, k'=1, \ldots, p
    \end{cases}.
\label{eq:init}
\end{equation}
\\
The exact orthogonality between student nodes is a simplification for the analytical treatment, but numerics show that the qualitative dynamical behavior remains unchanged when small initial student-student overlaps are also introduced.

\section{Results}

%\textit{In this chapter, we present the main results of our work.\\}

To study the system's evolution, one has to solve the coupled ODE system in Equation~\eqref{eq:ode_system}. While explicit analytical solutions are generally unavailable, numerical integration allows exploration of the dynamics across different initial conditions and learning rates $\eta$. As a sanity check, we compare the numerical solution of the ODEs with the dynamics observed in finite-$N$ simulations, using matched initialization conditions. In the simulations, we randomly construct two independent sets of orthonormal vectors: $p$ vectors for the student perceptrons and $p^*$ vectors for the teacher perceptrons. The initial teacher-student overlap matrix $\bm{\rho}$ emerges naturally from finite-$N$ fluctuations, distributed as $\mathcal{N}(0,1/N)$. For the dynamics, we apply the one-pass SGD weight update defined in Equation~\eqref{eqn:SGD}. We initialize the ODE system with the same overlap matrix $\bm{\rho}$ to ensure a consistent comparison.

\begin{figure}[h]
    \centering

    % First row
    \begin{subfigure}[t]{0.45\linewidth}
        \centering
        \includegraphics[width=\linewidth]{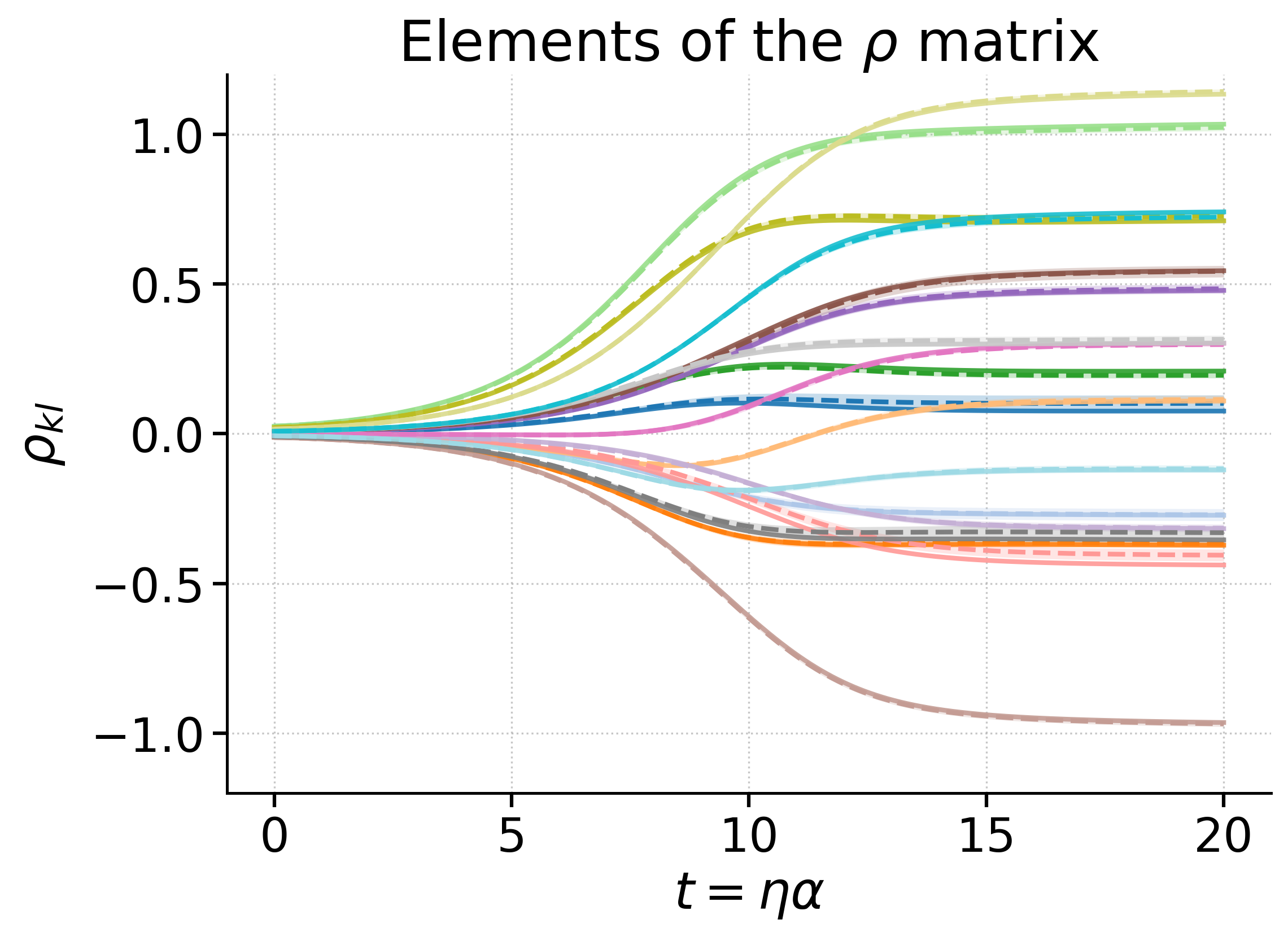}
    \end{subfigure}
    \hspace{0.05\linewidth}
    \begin{subfigure}[t]{0.45\linewidth}
        \centering
        \includegraphics[width=\linewidth]{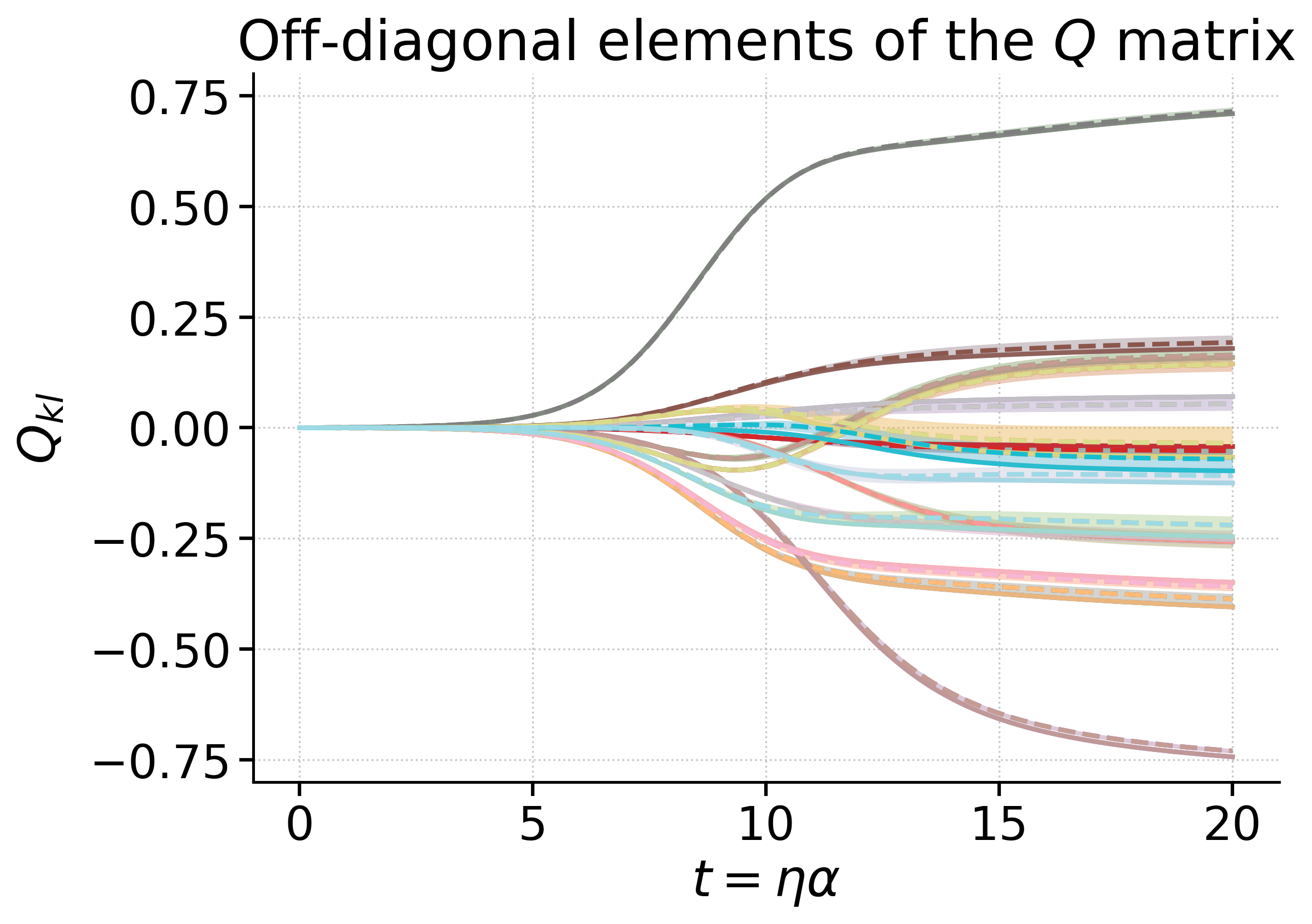}
    \end{subfigure}

    \vspace{1em}

    % Second row
    \begin{subfigure}[t]{0.45\linewidth}
        \centering
        \includegraphics[width=\linewidth]{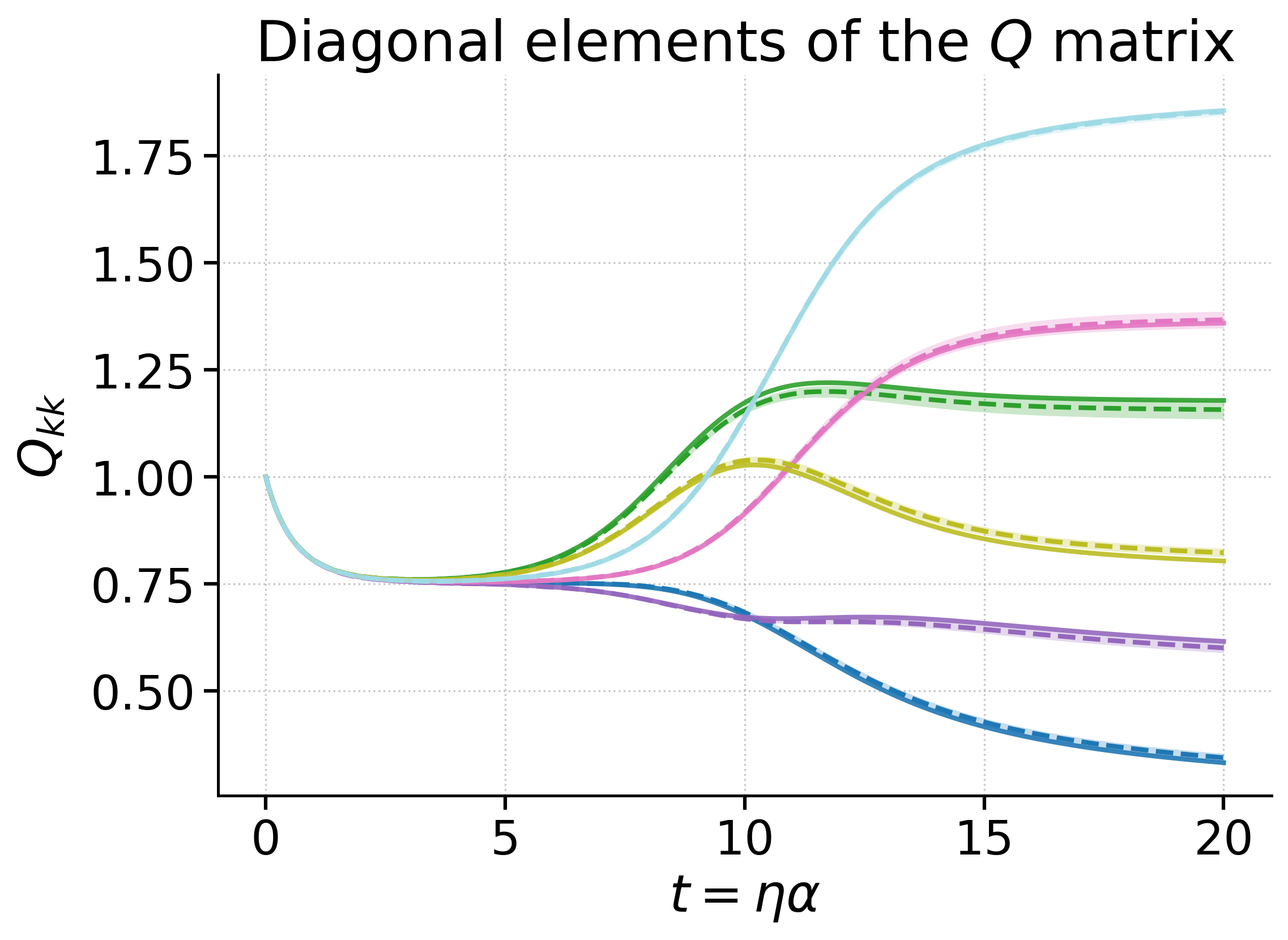}
    \end{subfigure}
    \hspace{0.05\linewidth}
    \begin{subfigure}[t]{0.45\linewidth}
        \centering
        \includegraphics[width=\linewidth]{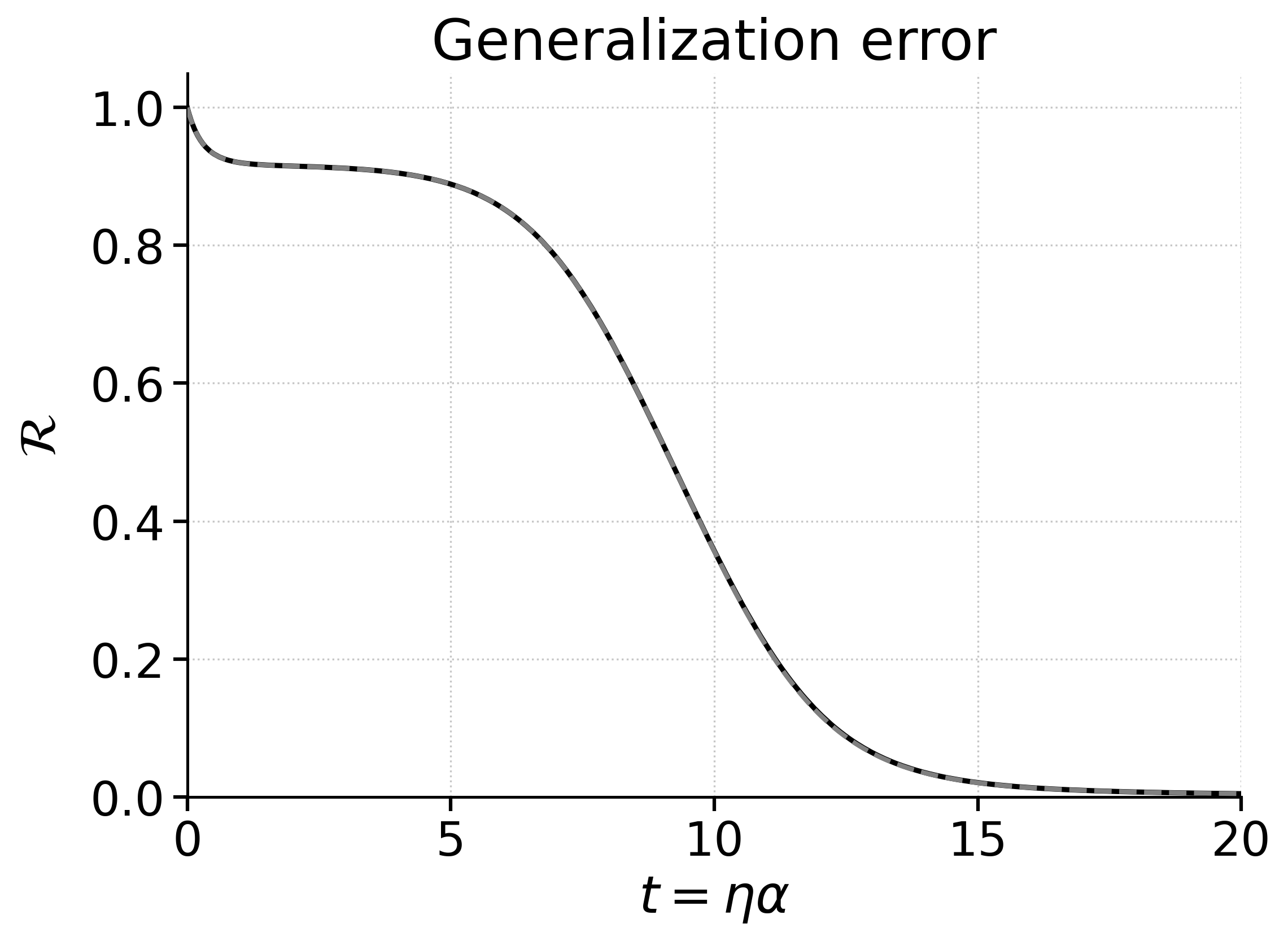}
    \end{subfigure}

    \caption{Comparison between the numerical solution of the ODEs (solid lines) and the average over 10 simulated dynamics (dashed lines), all initialized with the same random overlap configuration, for $p = 6$, $p^{*} = 3$, $\eta = 0.01$, and $N = 10^4$. The shaded region indicates the standard error of the mean for the simulations. \textit{Top left:} teacher-student overlaps. \textit{Top right:} student-student overlaps. \textit{Bottom left:} student norms. \textit{Bottom right:} generalization errors.}
    \label{fig:simulation_comparison}
\end{figure}

Figure~\ref{fig:simulation_comparison} provides a visual comparison between the theoretical and simulation-based evolution of all order parameters in a scenario where the student and teacher networks have different numbers of nodes (specifically, $p = 6$ and $p^* = 3$). The figure shows that the curves obtained from the ODE solutions closely align with the behavior observed in simulations, validating the theoretical predictions for the network’s dynamics.

To derive analytical results, we focus on the case of a very small learning rate $\eta \ll 1$. In this regime, the dynamics become effectively independent of $\eta$ (apart from a rescaling of the time scale) and reduce to a gradient flow in the population risk landscape \cite{veiga2022phase}. Therefore, it is useful to define a rescaled time variable, $t = \eta \alpha$, which removes the explicit dependence on $\eta$ in the equations. 
% The effects of a finite learning rate will be addressed at the end of Section \ref{sec:different_regimes}.

The evolution of the student network is characterized by distinct phases in the learning process, with qualitatively distinct behaviors, each corresponding to different timescales in  $t$. %These phases exhibit qualitatively different behaviors in the order parameters, which reflect transitions in the learning dynamics. 

In Section \ref{sec:different_regimes}, we explore these phases in detail and derive their key features directly from the ODEs. We then evaluate the time required for the student network to achieve signal recovery, and relate this timescale to the number of neurons in the student network, $p$. Finally, we characterize the configurations of the student network that result in zero generalization error, explain how these configurations can be constructed, and describe their properties.

In Section \ref{sec:landscape}, we analyze the population risk landscape $\mathcal{R}(w)$, identifying the critical points and evaluating the eigenvalues of the Hessian matrix at these points. We then connect these findings to the dynamics of the network, demonstrating the consistency between the predictions from both approaches.

\subsection{Different regimes during learning}
\label{sec:different_regimes}

\paragraph{Norm learning} In the initial phase of the dynamics, the norms of the student perceptrons, $Q_{kk}$, evolve significantly, while the overlaps between perceptron pairs, both student-student and teacher-student, remain negligible. This phase leads to a reduction in the population risk without requiring alignment between the student and teacher networks. In other works \cite{arnaboldi2023escaping}, this phase is bypassed by fixing the norms of the student perceptrons to their initial values.
As anticipated, we assume small variance $\epsilon^2$ for the initialization of $\rho_{kl}$ and $Q_{k,k'}$ with $k\neq k'$ in the ODEs (see Eq. \eqref{eq:init}), mimicking the small random fluctuations of finite-size simulations around the completely uninformed initialization. In this regime, the evolution of the order parameters is governed by the following system of equations:

\begin{equation}
  \begin{cases}
    \displaystyle
    \frac{d\rho_{kl}}{dt}
    = \frac{4}{p}\,\rho_{kl}
      \Biggl[1 + \frac{2}{p^{*}}
            - \frac{1}{p}\left(\sum_{r=1}^{p}Q_{rr} + 2Q_{kk}\right)
      \Biggr]
    + \mathcal{O}(\epsilon^{2}),
    &\forall k, l, \\[1em]
    
    \displaystyle
    \begin{aligned}[t]
      \frac{dQ_{kk'}}{dt}
      &= \frac{8}{p}\,Q_{kk'}
         \Biggl[
           1 - \frac{1}{p}\left(\sum_{r=1}^{p}Q_{rr} + Q_{kk} + Q_{k'k'}\right)
         \Biggr] \\
      &\quad - \frac{8}{p^{2}}\,Q_{kk'}\,(Q_{kk} + Q_{k'k'})
        + \mathcal{O}(\epsilon^{2}),
    \end{aligned}
    &\forall k \neq k', \\[1em]
    
    \displaystyle
    \frac{dQ_{kk}}{dt}
    = \frac{8}{p}\,Q_{kk}
      \Biggl[1 - \frac{1}{p}\left(\sum_{r=1}^{p}Q_{rr} + 2Q_{kk}\right)\Biggr]
    + \mathcal{O}(\epsilon^{2}),
    &\forall k.
  \end{cases}
  \label{eq:normLearningOdes}
\end{equation}

As long as the student-student overlaps ($Q_{kk'}$ with $k \neq k'$) and the student-teacher overlaps ($\rho_{kl}$) remain of order $\mathcal{O}(\epsilon)$, and assuming a sufficiently small learning rate $\eta$, their time derivatives also remain of order $\mathcal{O}(\epsilon)$. In contrast, the diagonal norm terms $Q_{kk}$ are initialized to $1$, and their derivatives are of order $\mathcal{O}(1)$, allowing these norms to evolve on a much faster timescale.
Moreover in this regime each equation for $Q_{kk}$ depends predominantly on its own value and the norms of the other student nodes, but not on their overlaps. As a result, the norms evolve independently of directional alignment, and  converge, 
on a time scale of order $\mathcal{O}(1)$, to an attractive fixed point where all student norms are equal to a $\bar{Q}$ defined as 

\begin{equation}
    \bar{Q} = \frac{p}{p + 2}.
\end{equation}

%This value represents an attractive fixed point for the norms. Since the norm derivatives are of order $\mathcal{O}(1)$, convergence to this fixed point occurs on a timescale significantly shorter than that of the overlap parameters, which remain small and contributeminimally to the dynamics of norm adjustment.

\paragraph{The plateau}
After the norms adjust to their fixed point, the learning dynamics enters a phase in which the reduction in population risk slows significantly, as shown in the bottom right panel of Figure~\ref{fig:simulation_comparison}. In this regime, the primary evolution occurs in the overlap terms, while the norms remain close to the fixed value $\bar{Q}$. As discussed in Section~\ref{sec:landscape}, this phase corresponds to a region of the population risk landscape characterized by numerous flat directions, which tend to keep the dynamics around what is often referred to as the \textit{plateau}. Escaping this region marks the onset of actual learning, where the agreement between the student and teacher networks macroscopically increases. In this region, we can rewrite the ODEs system in Equation~\eqref{eq:normLearningOdes}, substituting $\bar{Q}$ to the student nodes norms, as

\begin{equation}
    \begin{cases}
      \displaystyle\frac{d\rho_{kl}}{dt} = \omega_{\rho}(p, p^{*})\, \rho_{kl} + \mathcal{O}(\epsilon^{2}),
      &\forall\, k \in [1, p],\ l \in [1, p^{*}], \\[1em]
      
      \displaystyle \frac{dQ_{kk'}}{dt} = \omega_{Q}(p)\, Q_{kk'} + \mathcal{O}(\epsilon^{2}),
      &\forall\, k \neq k',\ k, k' \in [1, p], \\[1em]
      
      \displaystyle \frac{dQ_{kk}}{dt} = \mathcal{O}(\epsilon^{2}),
      &\forall\, k \in [1, p],
    \end{cases}
    \label{eq:plateau_odes}
\end{equation}

where the coefficients are defined as:
\begin{equation}
    \omega_{\rho}(p, p^{*}) = \frac{8}{p p^{*}}, \qquad \omega_{Q}(p) = \frac{16}{p(p+2)}.
\end{equation}
Focusing on the leading-order terms, we observe a set of decoupled equations in which the derivatives of the overlap parameters are proportional to the parameters themselves. This structure leads to a simple exponential growth in the overlap terms. Although the student perceptrons begin to align with the teacher, each perceptron evolves independently, with no interaction with the others. This independence persists until the overlaps reach finite values. Importantly, the absence of inter-perceptron interactions implies that \emph{overparameterization does not significantly help in escaping the plateau phase}. The only advantage is given by the fact that more than one node is independently trying to align and, as we will see, the global time scale for reconstruction is controlled by the node that, by chance, had the largest alignment in the initial condition.  \\

 \paragraph{Time of escape} We now turn to the problem of estimating the average time required for the student network to escape from the plateau region. This escape time essentially measures how long it takes the network to begin retrieving the signal, since the final descent toward the solution manifold occurs on a much shorter timescale. To estimate this escape time, we assume that on such time scale the order parameters remain sufficiently small such that the ODEs in Equation~\eqref{eq:plateau_odes} remain valid throughout.

To better understand the role of overparameterization, we want to compare results across different student network sizes $p$. A key question arises when making this comparison: should we use the same (small) constant learning rate $\eta$ for all values of $p$, or should $\eta$ scale with the network size? To answer this, we examine how the optimal learning rate, defined as the rate that yields the fastest convergence, scales with $p$.

\begin{figure}[h!]
    \centering
    \begin{subfigure}[b]{0.49\linewidth}
        \centering
        \includegraphics[width=\linewidth]{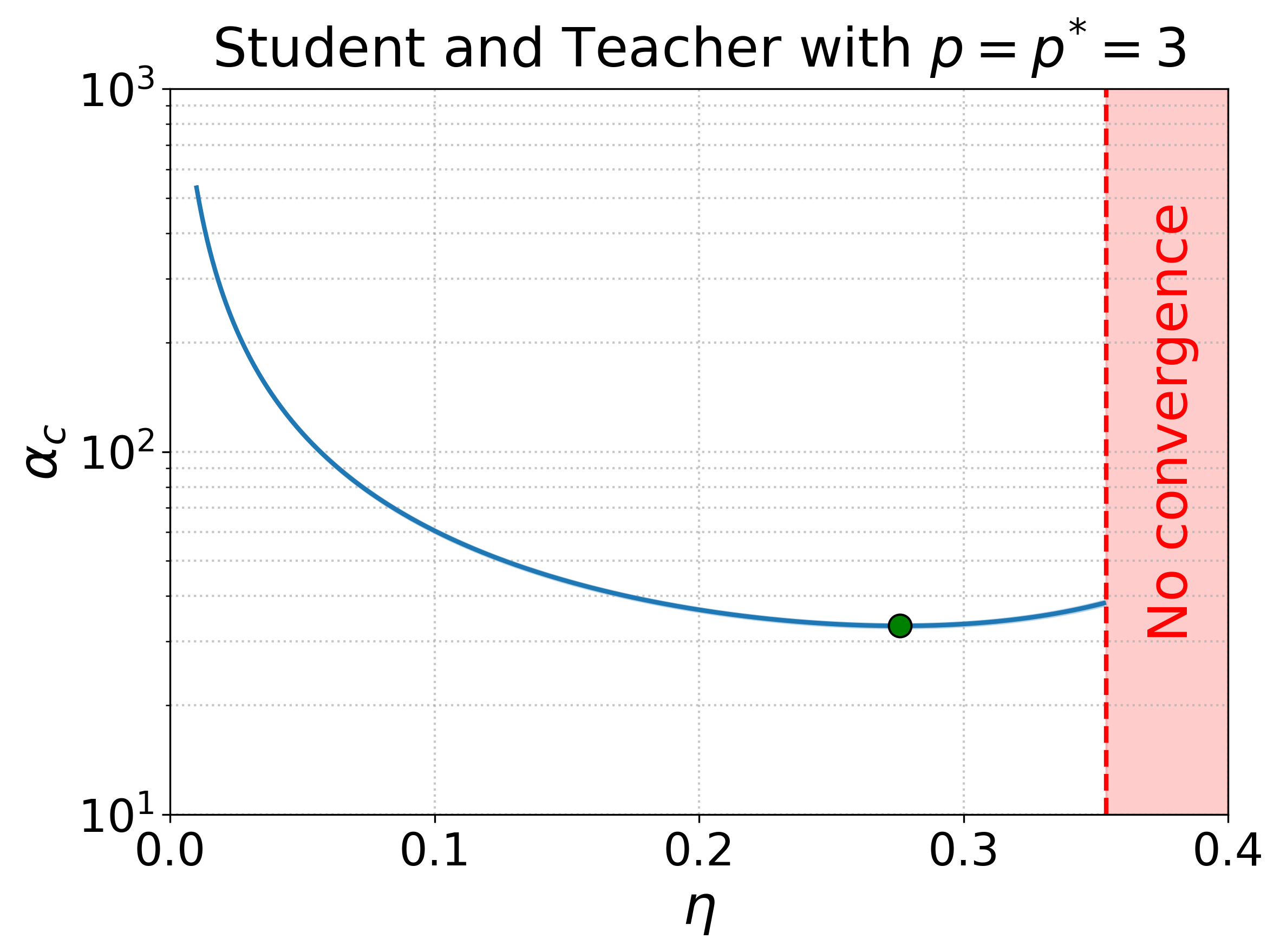}
        \label{fig:learning_rate_scaling_1}
    \end{subfigure}
    \hfill
    \begin{subfigure}[b]{0.49\linewidth}
        \centering
        \includegraphics[width=\linewidth]{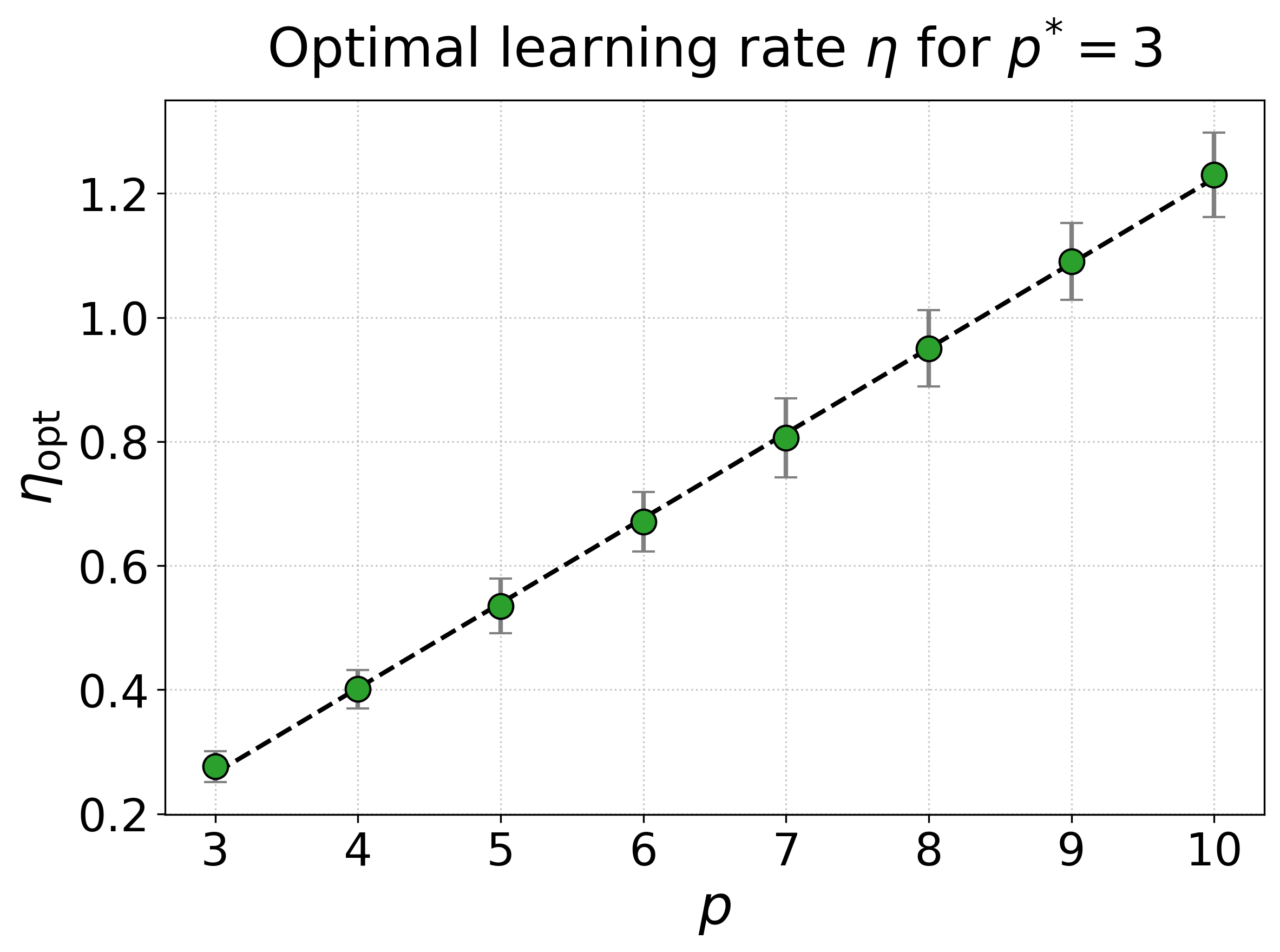}
        \label{fig:learning_rate_scaling_2}
    \end{subfigure}
    \caption{Numerical analysis of the optimal learning rate $\eta$ for student-teacher training dynamics. 
    \textit{Left}: Value of $\alpha_{c}$ at which the loss is reduced to half of its initial value, shown as a function of $\eta$ for fixed $p = p^* = 3, \,\epsilon=10^{-2}$. Each point of the curve corresponds to the average over 20 different random initializations.
    \textit{Right}: The optimal learning rate as a function of student width $p$, revealing a linear scaling trend.}
    \label{fig:learning_rate_scaling}
\end{figure}

As already discussed, analytical study of the dynamical phases is only possible under the assumption of a small learning rate $\eta$, which in this acts merely as a multiplicative scaling factor in the ODEs. Consequently, the scaling of the optimal learning rate $\eta_{\text{opt}}$ cannot be derived analytically from these equations alone. Instead, we adopt a heuristic mixed approach: for each combination of $p$ and $p^{*}$, we solve the corresponding ODEs numerically across a range of learning rates $\eta$. For each value of the learning rate we average over a random set of different initial conditions. Starting from small values, we increase $\eta$ until the system fails to converge, identifying the learning rate that minimizes the amount of data $\alpha_{c}$ required to escape the plateau. This provides a numerical estimate for $\eta_{\text{opt}}(p)$. Results for $p^{*} = 3$ are shown in Figure~\ref{fig:learning_rate_scaling}.

Our results indicate that $\eta_{\text{opt}}(p)$ scales linearly with $p$. In Equation~\eqref{eq:update_amplitude}, the learning amplitude is scaled by a factor of $1/p$; choosing $\eta$ proportional to $p$ effectively removes this dependence, making the learning amplitude independent of the student network size. We adopt this scaling in our analysis of the average escape time from the plateau region.

Under this scaling, we analyze the plateau dynamics using the linearized ODEs and estimate the time required for the system to escape. By setting $t = 0$ at the onset of the plateau and averaging over Gaussian initial conditions for the overlaps, $\mathcal{N}(0, \epsilon^{2})$, we find that the average relative reduction of the population loss is
\begin{equation}
    \left\langle\frac{\mathcal{R}(0) - \mathcal{R}(t)}{\mathcal{R}(0)}\right\rangle = 4\epsilon^2
    \frac{\exp\left(16 t / p^{*}\right) -1}{
     \left(1 + 2 / p^{*} - p / (p + 2)\right)}.
\end{equation}

We observe that the relative loss reduction increases with $p$ due to the prefactor, while $p$ does not appear in the characteristic timescale of the exponential, which is fixed to $p^{*}/16$. 
Therefore, the time to escape from the plateau is largely governed by teacher complexity $p^{*}$ and is only weakly influenced by student width $p$.
%This implies that overparameterization
% while overparameterization leads to slightly larger overlaps with the teachers at early times, it 
%does not accelerate the exponential rate of escape. 

\paragraph{Zero Error Region}

When the dataset size becomes sufficiently large (equivalently, after many weight updates in the online learning), the student-teacher overlaps reach finite values, enabling the system to escape the plateau region and achieve rapid loss decay. In the realizable regime ($p \geq p^*$) with noiseless training data, the dynamics converge to configurations with zero population risk - fixed points of the online learning process.
A key distinction emerges between our framework and most classical toy cases, {\it e.g.} phase retrieval ($p^* = 1$) in the class of quadratic networks, where only a finite set of isolated solutions appear.  Our setup instead, possesses a continuous manifold of zero-error solutions in the student weight space. This can be seen by noticing that the student output in Equation~\eqref{eq:studentOutput} can be rewritten as:

\begin{equation}
    \hat{y}(\vec{x}) = \frac{1}{pN}\vec{x}^{\,T}\bm{W}^T\bm{W}\vec{x}
\end{equation}
where the rows of $\bm{W}\in \mathbb{R}^{p \times N}$ are the weights of the student perceptrons, i.e. $\bm{W}_{ki} = (\vec{w}_{k})_i$. 
This means that, for fixed $p$ and $N$, any two student matrices $\bm{W}_1$ and $\bm{W}_2$ with $\bm{W}_1^T\bm{W}_1 = \bm{W}_2^T\bm{W}_2$ produce identical outputs. This equivalence class includes all rotations $\bm{W}' = \bm{R}\bm{W}$ where $\bm{R} \in \mathbb{\bm{R}}^{p \times p}$ satisfies $\bm{R}^T\bm{R} = \mathbb{I}_p$. 
% This type of invariance has been extensively studied in the context of optimization over subspaces~\cite{edelman1998geometry}.
Using a similar argument for the teacher network, the zero-error condition—i.e., the requirement that the student and teacher networks produce identical outputs for all inputs—can be rewritten as
\begin{equation}
    \frac{\bm{W}^T\bm{W}}{p} = \frac{{\bm{W}^*}^T\bm{W}^*}{p^*}
\end{equation}
where the rows of $\bm{W}^* \in \mathbb{R}^{p^* \times N}$ are the weights of the teacher perceptrons, i.e. $\bm{W}^*_{ki} = (\vec{w}_{k}^*)_i$. 

The zero-error condition reveals two fundamental aspects of our framework. First, when $p^* > 1$, complete recovery of individual teacher perceptrons becomes impossible - only their equivalence class under rotation can be identified. Second, for $p \geq p^*$, the space of optimal solutions forms a continuous manifold parametrized by rectangular orthogonal matrices $\bm{S} \in \mathbb{R}^{p \times p^*}$ satisfying $\bm{S}^T\bm{S} = \mathbb{I}_{p^*}$. Each such matrix defines a valid solution through the transformation:

\begin{equation}
    \overline{\bm{W}}_S = \sqrt{\frac{p}{p^*}} \bm{S}\bm{W}^*.
    \label{eq:solutions_formula}
\end{equation}

We define the dimensionality of this solution space as the number of independent parameters needed to specify such rectangular matrices. For a $p \times p^*$ matrix $\bm{S}$ with orthonormal columns, there are initially $pp^*$ free parameters. The orthogonality constraints $\bm{S}^T\bm{S} = \mathbb{I}_{p^*}$ impose $p^*(p^* + 1)/2$ independent conditions, leaving exactly $pp^* - p^*(p^* + 1)/2$ degrees of freedom. Crucially, this dimension is positive whenever $p > 1$, meaning the solution space remains non-trivial even when student and teacher have equal widths ($p = p^*$). This degeneracy, intrinsic to the matrix structure, makes our multi-perceptron case qualitatively different from scenarios with a unique solution, where the dimensionality of the solution manifold is zero.\\

\textit{Which solution does the network converge to?} Having established the existence of multiple solutions in the student weight space, we now address how the dynamics select among them. Since the solution of the dynamical ODEs in Equation~\eqref{eq:ode_system} is deterministic, the final configuration depends entirely on the initial conditions.
We make the following claim, then prove it in the following section:
\begin{quote}
\textbf{Claim:} \emph{The student network converges to the zero-error solution that is closest to its starting point in Euclidean distance.} \end{quote}

\noindent To show this, let's consider arbitrary initial weights $W(0)$. Using the parameterization of zero-error solutions $\overline{\bm{W}}_{\bm{S}}$ (Eq.~\eqref{eq:solutions_formula}), we can express the square of the Euclidean distance between the initial condition and an arbitrary solution as a function of $\bm{S}$:
\begin{equation}
    D(\bm{S}) = d(\overline{\bm{W}}_{S}, \bm{W}(0)) = \sum_{k=1}^p \left\| \sqrt{\frac{p}{p^{*}}} \sum_{l=1}^{p^{*}} S_{kl} \vec{w}^{*}_l - \vec{w}_k(0) \right\|^{2}.
\end{equation}

To find the closest solution to the initial condition, we minimize this function with respect to $\bm{S}$ under the constraint $\bm{S}^T \bm{S} = \mathbb{I}_{p^{*}}$. By introducing Lagrange multipliers $\Lambda = \{\lambda_{l,l'}\}_{l,l' = 1}^{p^{*}}$, we enforce these orthogonality constraints and aim to minimize the function:

\begin{equation}
    \Tilde{D}(\bm{S}, \Lambda) = d(\overline{\bm{W}}_{S}, \bm{W}(0)) + \sum_{l,l'} \lambda_{l,l'} \left( \sum_{k} S_{k,l}S_{k,l'} - \delta_{l,l'} \right).
\end{equation}
From this minimization, we obtain a compact expression for the corresponding matrix $\overline{\bm{S}}$, which only depends on the initial overlaps between the student and teacher perceptrons:
\begin{equation}
    \Bar{\bm{S}}(\bm{\rho}_0) = \bm{\rho}_0\, \left[\bm{\rho}_0^T \bm{\rho}_0\right]^{-1/2},
    \label{eq:final_overlap}
\end{equation}
where the exponent $-1/2$ denotes the matrix inverse square root. From the matrix $\bm{S}$, we can immediately recover the corresponding zero-error solution in terms of the order parameters:
\begin{align}
    \overline\rho_{kl} &= \frac{\vec{w}_k \cdot \vec{w}^*_l}{N} = \sqrt{\frac{p}{p^*}}\sum_{l'=1}^{p^*}S_{kl'}\frac{\vec{w}^*_{l'}\cdot\vec{w}^*_l}{N}  \implies \overline{\bm{\rho}} = \sqrt{\frac{p}{p^*}}\bm{S}\\
    \overline{Q}_{kk'} &= \frac{\vec{w}_k \cdot \vec{w}_{k'}}{N} = \frac{p}{p^*}\sum_{l,l'=1}^{p^*}S_{kl'}S_{k'l'}\frac{\vec{w}^*_{l'}\cdot\vec{w}^*_l}{N} \implies \overline{\bm{Q}} = \frac{p}{p^*}\bm{S}\bm{S}^T
\end{align}
where we have used that $(\vec{w}^*_{l}\cdot\vec{w}^*_{l'}) = N\delta_{l,l'}$. 
Since the noiseless dynamics are completely determined by the initial conditions, every point along a given trajectory can itself be viewed as a new initialization that would lead to the same final solution. One can indeed show that the quantity $\overline{\bm{S}}(t)$, defined by:
\begin{equation}
\overline{\bm{S}}(t) = \bm{\rho}(t) \left[\bm{\rho}(t)^T \bm{\rho}(t)\right]^{-1/2},
\label{eqn:final_overlap_const}
\end{equation}
remains constant throughout the dynamics (see Section \ref{app:S_conservation} of the Appendix for details). 

\begin{figure}[htbp]
    \centering
    \begin{minipage}{0.48\textwidth}
        \centering
    \includegraphics[width=\linewidth]{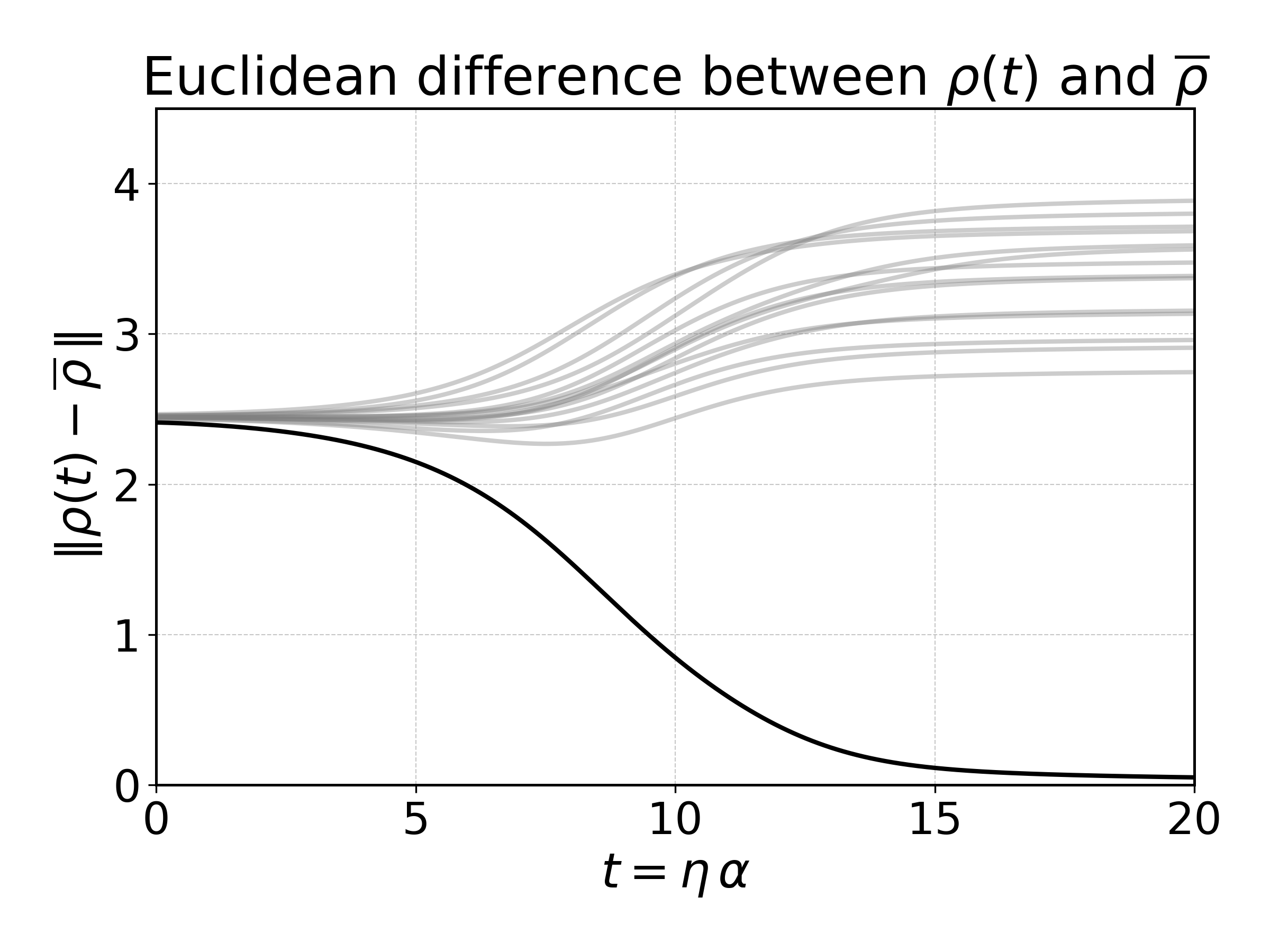}
    \end{minipage}
    \hfill
    \begin{minipage}{0.48\textwidth}
        \centering
        \includegraphics[width=\linewidth]{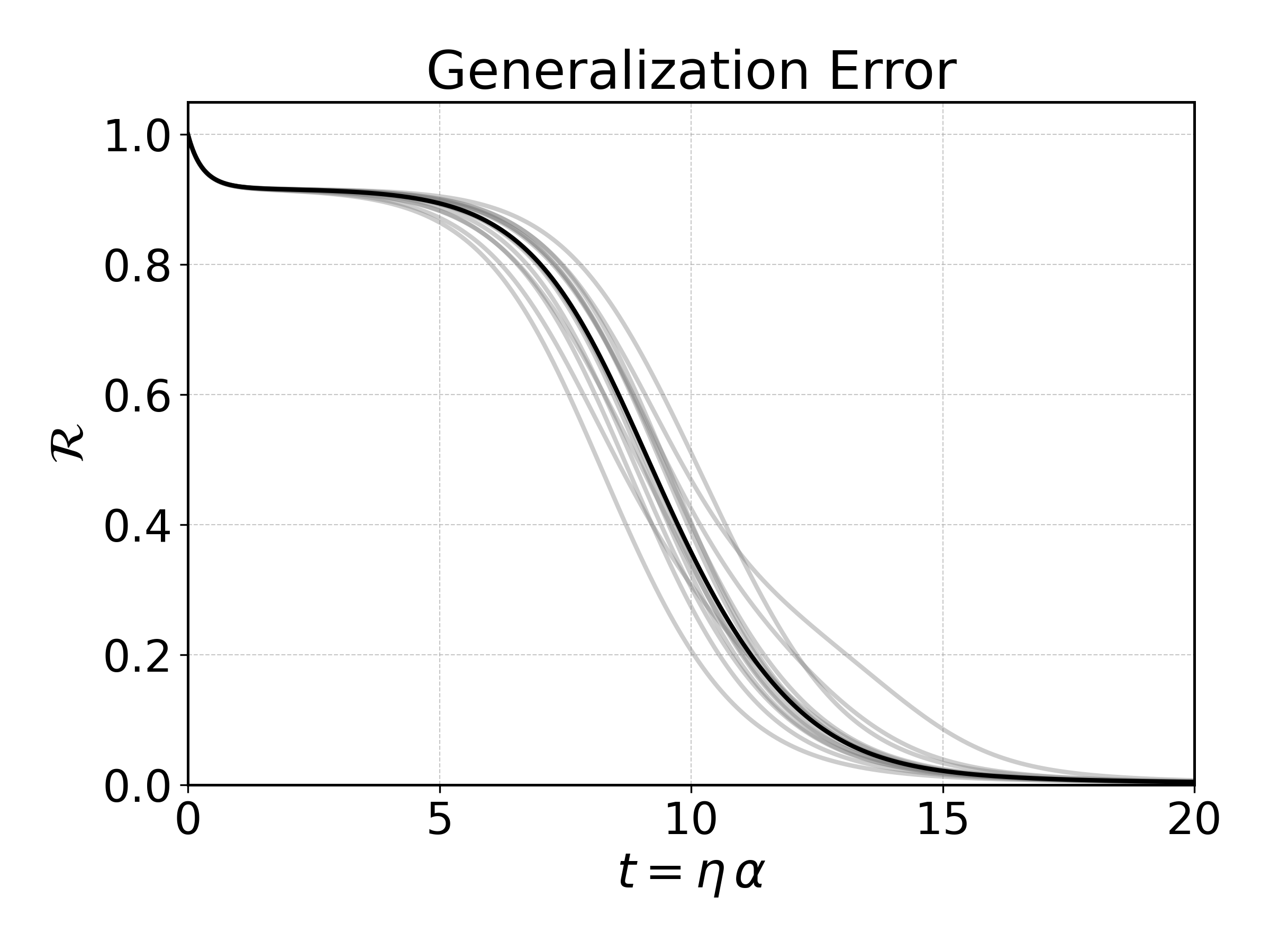}
    \end{minipage}
    \caption{
        Analysis using the numerical solution of the ODEs for $p = 6$, $p^{*} = 3$, $\eta = 0.01$, $\epsilon=0.01$.
        \textit{Left:} Euclidean distance between the evolving order parameter $\bm{\rho}(t)$ and the predicted zero-error solution $\overline{\bm{\rho}}$, shown in black for the trajectory initialized with $\bm{\rho}_0$ (used to compute $\overline{\bm{\rho}}$), and in gray for several trajectories initialized with uncorrelated random matrices.
    \textit{Right:} Evolution of the corresponding generalization error for all initializations.
    }
\label{fig:overlap_comparison}
\end{figure}

To verify our prediction, in Figure~\ref{fig:overlap_comparison}  we show the evolution of Euclidean distance with a given initial $\bm{\rho}_0$ and the corresponding prediction given by Equation~\eqref{eq:final_overlap} (black curve of Left Panel). We compare this with the Euclidean distance between this final zero-error point and dynamics with other initializations (grey lines). 
Note that all these dynamics converge to zero-error points as shown in the Right Panel.

The results confirm that Equation~\eqref{eq:final_overlap} accurately predicts the final solution when the initialization matches $\bm{\rho}_0$. For different initial conditions drawn from the same distribution, the system still converges to valid zero-error solutions, but these end points maintain a finite distance from the solution predicted for $\bm{\rho}_0$. 

% \paragraph{Effects of a finite learning rate}
% \todo{Questo paragrafo ancora non mi è chiaro dal punto di vista teorico. Dal punto di vista delle ODE è chiaro cosa cambia con learning rate finito: qualitativamente i tre regimi descritti sopra sono rispettati ma i valori predetti, ad esempio per il plateau, non sono più esatti. Il problema è dal punto di vista accordo numerica-ODE: facendo i test per N=10000, p=6 e pstar=3 viene fuori che per eta=0.01 le simulazioni (con la stessa inizializzazione) già variano leggermente tra di loro, ma sono abbastanza simili e le ODE sono una buona descrizione. per eta=0.05 e eta=0.1 le cose vanno sempre peggio e le dinamiche cambiano l'una dall'altra; La mia idea è che il miglior modo di capirlo è confrontare lo scaling della condizione iniziale (1/sqrt(N)) con la fluttuazione data da N-finito e vedere quando la dinamica delle simulazioni si dimentica della condizione iniziali.}

\subsection{Landscape}
\label{sec:landscape}

As established previously, for small learning rates the one-pass SGD dynamics approximate a gradient flow in the population risk landscape, while finite learning rates correspond to its discretized counterpart. This connection motivates our study of the risk landscape geometry, particularly its critical points and Hessian spectrum, to understand the learning dynamics. 
The population risk, representing the expected loss over the input distribution, admits the following expression in terms of order parameters:
\begin{equation}
\mathcal{R} = \frac{p^{*} + 2}{p^{*}} - \frac{2}{p} \sum_{k=1}^{p} Q_{k k} - \frac{4}{p p^{*}} \sum_{k,l} \rho_{k l}^{2} + \frac{1}{p^2}\left[3\sum_{k} Q_{kk}^2 + 2 \sum_{k < k'} \left(Q_{kk}Q_{k'k'} + 2Q_{kk'}^2 \right) \right].
\end{equation}

From this, we derive the gradient and Hessian with respect to student weights:
\begin{align}
\mathcal{G}_{m}&= \nabla_{\vec{w}_m}\mathcal{R} = -\frac{4}{p}\left[\vec{w}_m + \frac{2}{p^{*}}\sum_{l}\rho_{ml}\vec{w}^{*}_{l} - \frac{1}{p}\sum_{k} \left(Q_{kk}\vec{w}_{m} + 2Q_{km}\vec{w}_{k}\right) \right], \\
\mathcal{H}_{mn}(\Sigma) &= \nabla_{\vec{w}_m}\nabla_{\vec{w}_n}\mathcal{R} = -\frac{4}{p}\Bigg\{-\frac{2}{p}\left(Q_{nm}\mathbb{I}_{N} + \vec{w}_{n}\vec{w}_{m}^{T} + \vec{w}_{m}\vec{w}_{n}^{T}\right) \nonumber \\
&\quad + \delta_{mn}\left[\left(1 - \frac{1}{p}\sum_{k}Q_{kk}\right)\mathbb{I}_{N} + \frac{2}{p^{*}}\sum_{l}\vec{w}^{*}_{l}\vec{w}^{*T}_{l} - \frac{2}{p}\sum_{k}\vec{w}_{k}\vec{w}_{k}^{T} \right] \Bigg\}.
\end{align}

The full gradient $\vec{\mathcal{G}} \in \mathbb{R}^{pN}$ combines all component gradients $\mathcal{G}_m$ into a single vector, and similarly the full Hessian $\bm{\mathcal{H}} \in \mathbb{R}^{pN \times pN}$ organizes all $\mathcal{H}_{mn}$ blocks into a unique matrix. Using the compact matrix notation where $\bm{W} = [\vec{w}_1 \cdots \vec{w}_p]^T \in \mathbb{R}^{p \times N}$ represents the student weights and $\bm{W}^* = [\vec{w}^*_1 \cdots \vec{w}^*_{p^*}]^T \in \mathbb{R}^{p^* \times N}$ represents the teacher weights, the critical point condition $\mathcal{G}_m = \vec{0}$ for all $m$ reveals three fundamental classes of stationary points in the landscape. These distinct categories of stationary points correspond to different configurations of the student network relative to the teacher, each with particular implications for the learning dynamics.

\paragraph{Tabula rasa} The simplest critical point corresponds to the tabula rasa initialization where all student weights vanish $\bm{W} = \mathbf{0}$. This configuration yields order parameters:
\begin{equation}
    Q_{kk'} = 0, \;  \rho_{kl} = 0 \quad \forall k,k' \in [1,p], l \in [1,p^{*}]
\end{equation}
A complete analysis of the Hessian at this point reveals all eigenvalues are negative, confirming this trivial solution represents a local maximum in the risk landscape. This aligns with intuition - starting from zero weights, any small perturbation will decrease the population risk during training.

\paragraph{Student Networks Uncorrelated with Teacher} A more interesting class of stationary points occurs when student weights are non-zero but remain orthogonal to all teacher perceptrons ($\rho_{kl} = 0$ for all $k,l$), corresponding to complete failure of signal recovery. These configurations satisfy the orthogonality condition:
\begin{equation}
\bm{W} \bm{W}^{*T} = \mathbf{0}.
\end{equation}
For any rank $r \in [0,p]$, stationary solutions exist where the student weight matrix takes the form:
\begin{equation}
\bm{W} = \sqrt{\frac{p}{2 + r}} \bm{M} \bm{O}^T,
\end{equation}
where $\bm{M}$ is a $p \times r$ matrix with $\bm{M}\bm{M}^T = \mathbb{I}_r$, and $\bm{O}^T$ contains $r$ orthonormal vectors orthogonal to the teacher subspace. The most dynamically relevant case occurs when $r=p$, where student perceptrons are mutually orthogonal with quadratic norm $Q_{kk} = p/(p+2) \;\forall k$ and orthogonal to the teacher - precisely the plateau configuration described in Section \ref{sec:different_regimes}. Analysis of the Hessian reveals these points are saddles with mixed positive, negative, and null eigenvalues. 
The ratio of negative to null eigenvalues, which governs escape dynamics, is equal to:
\begin{equation}
\frac{\#\lambda_{<0}}{\#\lambda_{=0}}= \frac{p^*}{N - p^* - p/2 - 1/2} \ .
\end{equation}
This weak $p$-dependence (only appearing in subleading terms) is consistent with the decreasing with $p$ of the population loss as a function of time by only a prefactor, as found in Section \ref{sec:different_regimes}.

\paragraph{Student Networks Correlated with Teacher}
Another important class of stationary points emerges when student perceptrons develop meaningful correlations with the teacher network. For any rank $s \in [0,p^*]$, valid configurations exist where the student weight matrix combines teacher vectors according to:

\begin{equation}
\bm{W} = \sqrt{\frac{p(2 + p^*)}{p^*(2 + s)}} \bm{M} \bm{W}^{*T},
\end{equation}

where $\bm{M}$ represents a $p \times s$ matrix satisfying $\bm{M}\bm{M}^T = \mathbb{I}_s$. The dynamically relevant scenario occurs at full rank ($s=p^*$), where students completely span the teacher subspace, exactly matching the zero-error solutions previously identified in Eq.~\eqref{eq:solutions_formula}. The Hessian matrix at these optimal configurations exhibits a spectrum with only positive and null eigenvalues. There are $ p^*(N - p^*) + p^*(p^* + 1)/2$ positive eigenvalues, corresponding to directions that point away from the solutions manifold.

The null eigenvalues separate into two distinct categories. First, it is possible to identify $pp^* - p^*(p^*+1)/2$ directions corresponding to the linearization of the rotations, which generate the family of equivalent zero-error solutions via $\bm{S}$.
To give an intuition of their number we observe that it corresponds to the number of the degrees of freedom of the matrix $\bm{S}$ in Equation~\eqref{eq:solutions_formula}.
More specifically, in Appendix~\ref{sec:tangent_flat} we show that these $pp^* - p^*(p^*+1)/2$ eigenvectors correspond to the linear perturbations $\delta \bm{S}$ that preserve the zero-error constraint at first order, and therefore coincide with the tangent space of the manifold of zero-error solutions.
%directions tangent to the manifold of zero-error solutions, corresponding to the degrees of freedom of the matrix $\bm{S}$ in Equation~\eqref{eq:solutions_formula}. These are the flat directions connecting equivalent optimal configurations, obtained by linearizing the family of equivalent zero-error solutions generated by rotations of $\bm{S}$. In other words, they are the infinitesimal version of moving along the solution manifold without changing the represented function (see Appendix~\ref{sec:tangent_flat} for details). 
Second, there are $(N-p^*)(p-p^*)$ additional directions associated to null eigenvalues, which do not come from linearizing the continuous symmetry of the problem, and they are direct consequence of overparameterization as they only appear in the case $p>p^*$.

Notably, the persistence of flat directions even in the non-overparameterized case ($p=p^*$) reveals the fundamental role of rotational symmetry in this kind of multi-perceptron networks. This inherent symmetry guarantees a continuous solution manifold %regardless of the specific network dimensions 
as long as $p=p^*>1$.
More generally, for the same reason, it is in principle always possible to define continuous manifolds at constant loss regardless of the specific network dimensions as long as $p>1$, and therefore continuous manifolds of non zero minimal loss even in the underparametrized cases.
%explaining the observed learning dynamics where multiple equivalent solutions exist for any network configuration with $p>1$.

\section{Discussion}

We investigated the learning dynamics of an overparametrized student–teacher model with finite hidden widths $p$ and $p^*$ respectively, and with quadratic activation functions, trained via one-pass stochastic gradient descent. In the high-dimensional limit, the evolution of the system is governed by a set of deterministic ODEs that accurately capture the training dynamics. We focused on the small learning rate regime, corresponding to a gradient flow on the population loss dynamics. Solving these equations yields a comprehensive characterization of the learning process, which we complemented with an analysis of the geometry of the generalization risk landscape explored at different stages of the learning dynamics %interpreting the asymptotic configurations 
in terms of the structure of its critical points.

Our work shows that overparameterization only moderately accelerates the escape from the initial plateau of poor generalization, previously also observed in the 
\( p^* = 1 \)~\cite{arnaboldi2023escaping} case, without altering the characteristic timescale, which remains only determined by the difficulty of the problem encoded in the teacher complexity. 
The small impact of overparameterization during training is also evident in the landscape geometry of the portion of the landscape explored while trapped in the uninformative plateau region. In a situation in which the escape from the uninformative plateau is made difficult by the large number of flat directions in the landscape, overparameterization only introduces subleading (in the dimensionality of the data) terms that tend to decrease the fraction of null eigenvalues of the Hessian of typical critical points.  

The interest in the study of the case $p^* > 1$, beyond the simple generalization of results for
$p^* = 1$, becomes evident when we focus on the global minima. A central outcome of our analysis is the emergence of a continuous manifold of global minima in student–teacher models with quadratic activations when the teacher has more than one hidden unit $p^*>1$ and the student weights are unconstrained in norm. This degeneracy originates from a continuous rotational symmetry of the student weight matrix, which leaves the risk function invariant. Families of minimizers generated by symmetries in absence of overparameterization are less known in neural-network models, where invariances such as unit permutations or continuous rotations can produce degenerate solution manifolds \cite{zhao2022symmetries,zhao2025symmetry,cooper2021global, simsek2021geometry} in the overparametrized limit.
Remarkably, in our setting, flat directions associated with rotational invariance are already present at the interpolation point when $p=p^*>1$, showing that overparameterization is not required for marginal stability per se, but it can be seen as originated by the interplay among activation functions, the choice of the loss and the structure of the problem, contained in the teacher structure in our case and more generally encoded in the data set. Overparameterization nevertheless plays a crucial role in reshaping the geometry of the loss landscape. When 
$p>p^*$, we observe a proliferation of null eigenvalues in the Hessian at global minima, beyond those directly associated with symmetry transformations. This provides an explicit and analytically controlled example of how overparameterization favors increasingly wider minima, a mechanism often invoked in the literature also in connection with improved generalization \cite{draxler2018essentially, hochreiter1997flat, sagun2017empirical}.

Another important aspect is that, despite the resulting degeneracy of zero-loss solutions, learning dynamics does not converge to arbitrary points on the solution manifold. We find that stochastic gradient descent consistently selects the minimizer closest to initialization in Euclidean distance, revealing a clear form of implicit bias in gradient-based optimization.
%These phenomena are tightly connected to the growing literature on the implicit bias of gradient-based optimisation. In over-parametrized settings, gradient descent (and SGD) is known to pick, among the many global minimisers, those of minimal complexity: for instance, the max-margin separator in linear classification, or the minimum nuclear/path-norm solution in matrix and neural-network factorisations \cite{soudry2018implicit,gunasekar2017implicit,chizat2020implicit}. 
Similar solution-selection mechanisms have been identified in a variety of settings, %including linear models and matrix factorization, 
where gradient-based optimization favors geometrically distinguished solutions without explicit regularization: for instance, the max-margin separator in linear classification, or the minimum nuclear/path-norm solution in matrix and neural-network factorisations \cite{gunasekar2017implicit,gunasekar2018characterizing,soudry2018implicit,chizat2020implicit}.  % [Soudry et al., 2018; Gunasekar et al., 2018; Woodworth et al., 2020]. 
In our model, this behavior can be traced back to an explicit conservation law in the macroscopic dynamics which we identify and which constrains the accessible region of the solution manifold and uniquely determines the final state reached by SGD. 
%Indeed, as formalised by 
This result is in line with the 
Noether’s learning dynamics framework \cite{tanaka2021noether, zhao2022symmetries}:
when the loss exhibits continuous symmetries, the dynamics conserves the associated Noether charges, so the trajectory stays on the symmetry orbit fixed by the random initial weights, %gradient flow therefore cannot break these symmetries and 
merely sampling the degenerate manifold of solutions according to its starting point. %a fact formalised by Noether’s learning dynamics framework \cite{tanaka2021noether}. Together, these results underpin the “minimal-distance” principle we observe: the student converges to the closest point on the solution manifold that is compatible with both the data constraints and the conserved quantities imposed by the optimisation dynamics.

Furthermore, it is worth noting that the variance arising from the dependence of the learned weights on the initial conditions has recently been identified as one of the main contributions—together with the variance induced by dataset noise—to the overfitting peak at the interpolation threshold in the classical U-shaped curve associated with the bias–variance trade-off \cite{d2020double, geiger2020scaling}.
%[cit. Double Trouble in Double Descent : Bias and Variance(s) in the Lazy Regime and Geiger et al. 2019 citato nel precedente]. 
Within this framework, overparameterization plays a central role in the subsequent decrease of the generalization error, giving rise to the so-called double descent phenomenon: in highly overparameterized regimes, it effectively induces a form of self-averaging over the variability due to initial conditions.
Although our analysis of online learning directly minimizes the population risk landscape (not allowing us to distinguish between a train and test set), it nevertheless reveals how overparameterization generates an increasingly large set of marginal directions, corresponding to a broad manifold of zero-loss solutions, among which the final learned weights are selected in an initialization-dependent manner.
An interesting direction for future work would be to investigate the same teacher–student setup using more realistic finite-batch stochastic gradient descent on a finite dataset, where the generalization error is well defined. 
This more realistic setup should be at least partially reminiscent of the online setting and, if it allows one to explicitly observe double-descent behavior, it could further clarify the role of overparameterization in shaping generalization.

Finally, an additional interesting direction would be to explore possible connections with the specialization transition \cite{goldt2019generalisation, tian2020student},
%[arXiv:1909.13458, https://arxiv.org/pdf/1901.09085]
noting that in the presence of continuous symmetries in the space of solutions the usual definitions of specialization may not be strictly satisfied, even though the model converges to a functionally equivalent representation; this suggests that similar symmetry-induced ambiguities could also arise in more realistic settings, potentially challenging standard criteria used to detect specialization.

\section{Acknowledgments}

We thank Florent Krzakala for valuable discussions. We acknowledge financial support from the European Union – NextGenerationEU (PNRR) Fund, Mission 4, Component 2. In particular, we acknowledge support from Investment 1.1 – Project 202234LKBW "Land(e)scapes: Statistical Physics Theory and Algorithms for Inference and Learning Problems", CUP B53D23003850006 and CUP J53D23001330001 (PRIN 2022). This work was also supported by Investment 1.3 – FAIR Foundation, Extended Partnership “Future Artificial Intelligence Research” (Project Code PE00000013-FAIR)

%CL and CC acknowledge the European Union - Next Generation EU fund, component M4.C2, investment 1.1 - CUP J53D23001330001 (PRIN 2022). CC  also acknowledges the support of PNRR MUR project PE0000013-FAIR. 

%\cite{greenwade93}. %\href{https://www.overleaf.com/help/97-how-to-include-a-bibliography-using-bibtex}{video tutorial here}
%you can also import your Mendeley or Zotero library directly as a \verb|.bib| file, via the upload menu in the file-tree.

\bibliographystyle{ieeetr}
\bibliography{biblio}

\section{Appendix}

\subsection{Dynamical equations}
\label{sec:complete_equations}

In order to derive the equations that describe the dynamics of the order parameters in the main text, we start from
\begin{equation}
    \frac{d\rho_{kl}}{d\alpha}
    =
    \langle F_k u_l \rangle,
    \qquad
    \frac{dQ_{kk'}}{d\alpha}
    =
    \langle F_k \lambda_{k'} \rangle
    +
    \langle F_{k'} \lambda_k \rangle
    +
    \langle F_k F_{k'} \rangle,
    \label{eq:ode_system_app}
\end{equation}
where \(\eta\) is the learning rate and \(F_k^\mu\) is the learning amplitude,
\begin{equation}
    F_k^\mu
    =
    \frac{4\eta}{p}
    \left(
    \frac{1}{p^*}\sum_{l=1}^{p^*}(u_l^\mu)^2
    -
    \frac{1}{p}\sum_{r=1}^{p}(\lambda_r^\mu)^2
    \right)
    \lambda_k^\mu.
    \label{eq:update_amplitude_app}
\end{equation}

Since the variables \(u_l\) and \(\lambda_k\) are Gaussian, Wick's theorem allows us to explicitly expand the system \eqref{eq:ode_system_app}. Keeping all terms, one obtains for the student--teacher overlap
\begin{equation}
\frac{d\rho_{kl}}{d\alpha}
= \langle F_k u_l \rangle=
\frac{4\eta}{p}\rho_{kl}
\left[
1+\frac{2}{p^*}
-\frac{1}{p}
\left(
\sum_{r\ne k}^{p}Q_{rr}+3Q_{kk}
\right)
\right]
-\frac{8\eta}{p^2}\sum_{r\ne k}^{p}Q_{kr}\rho_{rl}.
\label{eq:rho_dynamics_full}
\end{equation}

For the student--student overlap (with diagonal terms corresponding to the norms), we obtain
\begin{align}
\frac{dQ_{kk'}}{d\alpha}
&=
\langle F_k \lambda_{k'} \rangle
+
\langle F_{k'} \lambda_k \rangle
+
\langle F_k F_{k'} \rangle
\\[0.5em]
&=
\frac{8\eta}{p}Q_{kk'}
\left[
1-\frac{1}{p}
\left(
\sum_{r=1}^{p}Q_{rr}+Q_{kk}+Q_{k'k'}
\right)
\right]
+
\frac{16\eta}{p\,p^*}\sum_{l=1}^{p^*}\rho_{kl}\rho_{k'l}
\nonumber\\
&\quad
-
\frac{8\eta}{p^2}\sum_{r\neq k}^{p}Q_{rk}Q_{rk'}
-
\frac{8\eta}{p^2}\sum_{r\neq k'}^{p}Q_{rk}Q_{rk'}
\nonumber\\[0.5em]
&\quad
+
\frac{16\eta^2}{p^2}
\Bigg\{
Q_{kk'}
\left[
1+\frac{2}{p^*}
-\frac{2p-1}{p^2}
\left(\sum_{r=1}^{p}Q_{rr}\right)^2
-
\frac{1}{p\,p^*}
\sum_{r=1}^{p}\sum_{l=1}^{p^*}\rho_{rl}^2
\right]
\nonumber\\
&\qquad
+
\frac{4}{p^*}\sum_{l=1}^{p^*}\rho_{kl}\rho_{k'l}
+
\frac{8}{(p^*)^2}\sum_{l=1}^{p^*}\rho_{kl}\rho_{k'l}
\nonumber\\
&\qquad
+
\frac{4}{p^2}
\left(\sum_{r=1}^{p}Q_{rr}\right)
\left(\sum_{r=1}^{p}Q_{kr}Q_{k'r}\right)
+
\frac{8}{p^2}\sum_{r_1,r_2=1}^{p}Q_{r_1r_2}Q_{kr_1}Q_{k'r_2}
\nonumber\\
&\qquad
-
\frac{4}{p}\sum_{r=1}^{p}Q_{kr}Q_{k'r}
-
\frac{4}{p\,p^*}
\left(\sum_{r=1}^{p}Q_{rr}\right)
\left(\sum_{l=1}^{p^*}\rho_{kl}\rho_{k'l}\right)
\nonumber\\
&\qquad
-
\frac{8}{p\,p^*}\sum_{r=1}^{p}\sum_{l=1}^{p^*}
\rho_{kl}\rho_{rl}Q_{k'r}
-
\frac{8}{p\,p^*}\sum_{r=1}^{p}\sum_{l=1}^{p^*}
\rho_{k'l}\rho_{rl}Q_{kr}
\Bigg\}.
\nonumber
\label{eq:Q_dynamics_full}
\end{align}

\subsection{Study of the population risk landscape}

We now study the teacher-student setup with quadratic activation using a different scaling as the one we usually use (the problem remains equivalent), as was also done and justified in \cite{arnaboldi2023escaping}. Unless otherwise stated, in the following text we will assume that $N \geq p \geq p^{*}$.
\\

\textbf{Teacher output given input $\vec{x}$:}
\[
y(\{\vec{w}^{*}\}, \vec{x}) = \frac{1}{p^{*}} \sum_{l=1}^{p^{*}} \left( \frac{\vec{\Tilde{T}}_{l} \cdot \vec{x}}{\sqrt{N}} \right)^2 = \frac{1}{p^{*}} \sum_{l=1}^{p^{*}} \left( \vec{w}^{*}_{l} \cdot \vec{x} \right)^2 = \frac{1}{p^{*}} \sum_{l=1}^{p^{*}} (u_{l})^2
\]

\textbf{Student output given input $\vec{x}$:}
\[
\hat{y}(\{\vec{w}\}, \vec{x}) = \frac{1}{p} \sum_{k=1}^{p} \left( \frac{\vec{\Tilde{J}}_{k} \cdot \vec{x}}{\sqrt{N}} \right)^2 = \frac{1}{p} \sum_{k=1}^{p} \left( \vec{w}_{k} \cdot \vec{x} \right)^2 = \frac{1}{p} \sum_{k=1}^{p} (\lambda_{k})^2
\]

\textbf{Scalings:}
\[
\left\{
\begin{array}{c}
\Tilde{T}_{l}^{i} \sim \Tilde{J}_{k}^{i} \sim \mathcal{N}(0, 1) \quad \Rightarrow \quad \|\vec{\Tilde{T}}_{l}\|^2 \sim \|\vec{\Tilde{J}}_{k}\|^2 \sim N \\[10pt]

w_{l}^{* i} \sim w_{k}^{i} \sim \mathcal{N}(0, 1/N) \quad \Rightarrow \quad \|\vec{w}^{*}_{i}\|^2 \sim \|\vec{w}_{k}\|^2 \sim 1 \\[10pt]

x^{i} \sim \mathcal{N}(0, 1) \quad \Rightarrow \quad \|\vec{x}\|^2 \sim N
\end{array}
\right.
\]
\\

Our goal is now to study the properties of the \textbf{loss landscape}. In order to do this, we can first write the loss function, the corresponding gradient and Hessian matrix as functions of the pre-activation outputs of the teacher ($u_{l} = \vec{w}^{*}_{l} \cdot \vec{x}$) and of the student ($\lambda_{k} = \vec{w}_{k} \cdot \vec{x}$).  \\

\textbf{Loss Function:}
\begin{align*}
\ell(\{u_{l}\}, \{\lambda_{k}\}, \vec{x}) &= (\hat{y}(\vec{w}, \vec{x}) - y(\vec{w}^{*}, \vec{x}))^2 = \left[ \frac{1}{p} \sum_{k=1}^{p} (\lambda_{k})^2 - \frac{1}{p^{*}} \sum_{l=1}^{p^{*}} (u_{l})^2 \right]^2 \\
&= \frac{1}{p^2} \sum_{k, k' = 1}^{p} \lambda_{k}^2 \lambda_{k'}^2 + \frac{1}{p^{*2}} \sum_{l, l' = 1}^{p^{*}} u_{l}^2 u_{l'}^2 
- \frac{2}{p p^{*}} \sum_{k=1}^{p} \sum_{l=1}^{p^{*}} \lambda_{k}^2 u_{l}^2 
\end{align*}

\textbf{Gradient:}
\begin{align*}
\nabla_{\vec{w}_p} \ell(\{u_{l}\}, \{\lambda_{k}\}, \vec{x}) &= (\nabla_{\vec{w}_p} \lambda_p) \frac{\partial}{\partial \lambda_p} \ell(\{u_{l}\}, \{\lambda_{k}\}, \vec{x}) \\
&= -\frac{4}{p}\left[\frac{1}{p^{*}}\sum_{l=1}^{p^{*}}u_{l}^{2} -\frac{1}{p}\sum_{k=1}^{p}\lambda_{k}^{2} \right] \lambda_{p} \vec{x}
\end{align*}

\textbf{Hessian:}
\begin{align*}
\nabla_{\vec{w}_{q}} \nabla_{\vec{w}_{p}} \ell(\{u_{l}\}, \{\lambda_{k}\}, \vec{x}) &= (\nabla_{\vec{w}_q} \lambda_q) \frac{\partial}{\partial \lambda_q} \nabla_{\vec{w}_p} \ell(\{u_{l}\}, \{\lambda_{k}\}, \vec{x}) \\
&= \frac{4}{p} \vec{x}\vec{x}^T \left\{ \frac{2}{p} \lambda_{q}\lambda_{p} - \delta_{qp} \left[\frac{1}{p^{*}}\sum_{l=1}^{p^{*}}u_{l}^{2} -\frac{1}{p}\sum_{k=1}^{p}\lambda_{k}^{2} \right] \right\}
\end{align*}

We can now average over the input distribution ($x^{i} \sim \mathcal{N}(0, 1)$, i.i.d.) to obtain the averaged version of the three quantities as functions of the order parameters. Note that this also corresponds to the study of the geometry of the \textbf{generalization risk}. We remind that the order parameters $\{\mathcal{Q}\}$ (using the new scalings) are defined as:
\begin{align*}
\begin{cases}
    Q_{k k'} &= \vec{w}_{k} \cdot \vec{w}_{k'} \\
    \rho_{k l} &= \vec{w}_{k} \cdot \vec{w}^{*}_{l}
\end{cases}
\end{align*}

\textbf{Generalization Risk:}
\begin{align*}
\mathcal{R}(\{\mathcal{Q}\}) &= \mathbb{E}_{x} \left[ \ell(\{u_{l}\}, \{\lambda_{k}\}, \vec{x}) \right] = \\
 &= \frac{p^{*} + 2}{p^{*}} - \frac{2}{p} \sum_{k=1}^{p} Q_{k k} - \frac{2}{p p^{*}} \sum_{k=1}^{p} \sum_{l=1}^{p^{*}} 2 \rho_{k l}^{2} + \frac{1}{p^2}\left[3\sum_{k=1}^{p} Q_{kk}^2 + 2 \sum_{k < k'}^{p} \left(Q_{kk}Q_{k'k'} + 2Q_{kk'}^2 \right) \right] 
\end{align*}

\textbf{Expected Gradient:}
\begin{align*}
\nabla_{\vec{w}_p} \mathcal{R}(\{\mathcal{Q}\}) &= \nabla_{\vec{w}_p} \mathbb{E}_{x}\left[\ell(\{u_{l}\}, \{\lambda_{k}\}, \vec{x}) \right] = \mathbb{E}_{x} \left[\nabla_{\vec{w}_p} \ell(\{u_{l}\}, \{\lambda_{k}\}, \vec{x}) \right] = \\
&= -\frac{4}{p}\left[\vec{w}_p + \frac{2}{p^{*}}\sum_{l=1}^{p^{*}}\rho_{pl}\vec{w}^{*}_{l} - \frac{1}{p}\sum_{k=1}^{p} \left(Q_{kk}\vec{w}_{p} + 2Q_{kp}\vec{w}_{k}\right) \right]
\end{align*}

\textbf{Expected Hessian:}
\begin{align*}
&\nabla_{\vec{w}_p} \nabla_{\vec{w}_q} \mathcal{R}(\{\mathcal{Q}\}) = \nabla_{\vec{w}_p} \nabla_{\vec{w}_q} \mathbb{E} \left[\ell(\{u_{l}\}, \{\lambda_{k}\}, \vec{x}) \right] = \mathbb{E} \left[\nabla_{\vec{w}_p} \nabla_{\vec{w}_q}\ell(\{u_{l}\}, \{\lambda_{k}\}, \vec{x}) \right] \\
&= -\frac{4}{p}\left\{-\frac{2}{p}\left(Q_{qp}\mathds{1}_{N} + \vec{w}_{q}\vec{w}_{p}^{T} + \vec{w}_{p}\vec{w}_{q}^{T}\right) + \delta_{qp}\left[\left(1 - \frac{1}{p}\sum_{k=1}^{p}Q_{kk}\right)\mathds{1}_{N} + \frac{2}{p^{*}}\sum_{l=1}^{p^{*}}\vec{w}^{*}_{l}\vec{w}^{*T}_{l} - \frac{2}{p}\sum_{k=1}^{p}\vec{w}_{k}\vec{w}_{k}^{T} \right] \right\}
\end{align*}

\subsubsection{Stationary points}

We can now look for stationary points in the generalization risk landscape by imposing the equation $\nabla_{\vec{w}_p} \mathcal{R}(\{\mathcal{Q}\}) = \vec{0}$.\\

It is useful to introduce the following matrices:
\begin{align*}
\begin{cases}
    W_{ki} &= \quad (\vec{w}_{k})_{i} \quad \text{\textit{$p \times N$ matrix}}\\
    W^{*}_{lj} &= \quad (\vec{w}^{*}_{l})_{j} \quad \text{\textit{$p^{*} \times N$ matrix}}\\
    \mathbb{G}_{\bm{W}} &= \quad \bm{W}^{T}\bm{W}\quad \text{\textit{$N \times N$ matrix}} \\
    \mathbb{G}_{\bm{W}^{*:}} &= \quad \bm{W}^{*T}\bm{W}^{*:} \quad \text{\textit{$N \times N$ matrix}} \\
\end{cases}
\end{align*}

We also notice that the order parameters of the model can be obtained from this matrices using the relations: 
\begin{align*}
\begin{cases}
    Q_{kk'} &= \quad \bm{W}\bm{W}^{T} \quad \text{\textit{$p \times p$ matrix}}\\
    \rho_{kl} &= \quad \bm{W}\bm{W}^{*T} \quad \text{\textit{$p \times p^{*}$ matrix}}
\end{cases}
\end{align*}

The stationarity equation $\nabla_{\vec{w}_p} \mathcal{R}(\{\mathcal{Q}\}) = \vec{0}$ can be rewritten in terms of these matrices as:
\begin{align*}
    \bm{W}+ \frac{2}{p^{*}}\bm{\rho}\bm{W}^{*:} - \frac{1}{p}\bm{W}\, \text{tr}(\bm{Q}) - \frac{2}{p}\bm{Q}\bm{W}&= \bm{0}\\
    \Rightarrow \mathbb{G}_{\bm{W}}\left[1 - \frac{1}{p}\, \text{tr}(\mathbb{G}_{\bm{W}})\right] - \frac{2}{p}\mathbb{G}_{\bm{W}}^2 + \frac{2}{p^{*}}\mathbb{G}_{\bm{W}}\mathbb{G}_{\bm{W}^{*:}} &= \bm{0}
\end{align*}

We can now study the second equation as a function of the matrices $\mathbb{G}_{W}$ and $\mathbb{G}_{\bm{W}^{*:}}$ only.

\paragraph{Trivial case: $\mathbb{G}_{\bm{W}} = \bm{0}$ ($\bm{W}= \bm{0}$) \\}  
This stationary point corresponds to the case where all the weights of the student are identical and equal to zero. We notice that this corresponds to the \textit{tabula rasa} initialization, showing how this cannot be used in the solution of the $N \to \infty$ equations.

\paragraph{Students uncorrelated to the teacher: $\mathbb{G}_{\bm{W}}\mathbb{G}_{\bm{W}^{*:}} = \bm{0}$ ($\bm{\rho}= \bm{0}$) \\}  

In this case the equation simplifies to:

\begin{equation*}
    \mathbb{G}_{\bm{W}}\left[1 - \frac{1}{p}\, \text{tr}(\mathbb{G}_{\bm{W}})\right] - \frac{2}{p}\mathbb{G}_{\bm{W}}^2 = \bm{0}
\end{equation*}

In order to have a solution, the matrix $\mathbb{G}_{\bm{W}}$ must be proportional to an orthogonal projection matrix, i.e.: 

\begin{equation*}
    \mathbb{G}_{\bm{W}} = \alpha \mathbb{P}
\end{equation*}

with $\mathbb{P}^{2} = \mathbb{P}$ and $\mathbb{G}_{\bm{W}}$ is symmetric by construction. Since we are considering the case with no correlation between student and teacher, the subspace onto which the matrix projects must be orthogonal to that formed by the teachers (i.e., all students must be orthogonal to the teachers). Since we know that for this type of matrices

\begin{equation*}
    \text{tr}(\mathbb{P}) = \text{rank}(\mathbb{P}) = \text{rank}(\bm{W}),
\end{equation*}

the equation reduces to finding $\alpha$ such that:

\begin{equation*}
    1 - \frac{1}{p}\text{rank}(\bm{W}) - \frac{2}{p}\alpha = 0 \iff \alpha = \frac{p}{2 + \text{rank}(\bm{W})}
\end{equation*}

This means that for every value of $r = \text{rank}(\bm{W}) \in [0, p]$ we have stationary points whenever the matrix $\bm{W}$ can be written as 
\begin{equation*}
    \bm{W}= \sqrt{\alpha} \mathbb{M} \mathbb{O}^{T}
\end{equation*}

where 

\begin{itemize}
    \item $\mathbb{M}$ is a generic $p \times r$ matrix such that $\mathbb{M}\mathbb{M}^T = \mathds{1}_{r}$
    \item $\mathbb{O^T}$ is a $r \times N$ matrix whose rows are orthonormal vectors ($\mathbb{O^T O} = \mathds{1}_{r}$) that are also orthogonal to the teacher vectors.
\end{itemize}
In terms of order parameters, these solutions correspond to:
\begin{align*}
\begin{cases}
    &\bm{Q} =\bm{W}\bm{W}^{T} = \alpha \mathbb{M} \mathbb{O}^T \mathbb{O} \mathbb{M}^T = \alpha \mathbb{M} \mathbb{M}^T\\
    &\bm{\rho} = \bm{0}
\end{cases}
\end{align*}

\paragraph{Example 1: all independent students ($r = \text{rank}(\bm{W}) = p$)\\}

In this case $\alpha = p/(2 + p)$, the matrices $\mathbb{M}$ are $p \times p$ dimensional and correspond to all possible rotation in $p$ dimensions. The overlap matrix is fully determined by:
\begin{equation*}
    \bm{Q} = \alpha \mathbb{M} \mathbb{M}^T = \alpha \mathds{1}_{p}
\end{equation*}

which corresponds to orthonormal students all with norm equal to $p/(2 + p)$.

\paragraph{Example 2: all students aligned ($r = \text{rank}(\bm{W}) = 1$)\\}

In this case $\alpha = p/3$, the matrix $\mathbb{M}$ corresponds to a generic $p$-dimensional vector $\bm{v}$ with norm one and the overlap matrix is equal to:

\begin{equation*}
    \bm{Q} = \alpha \mathbb{M} \mathbb{M}^T = \alpha \bm{v} \bm{v}^{T}
\end{equation*}

which corresponds to students all aligned to the same direction with quadratic norm $Q_{ii} = \alpha v_{i}^2$ such that $\sum_{k=1}^{p} Q_{ii} = \alpha \sum_{k=1}^{p} v_{i}^2 = \alpha = p/3$. 

\paragraph{Students correlated to the teacher: \\}  

In this case we have to consider the full equation 

\begin{equation*}
    \mathbb{G}_{\bm{W}}\left[1 - \frac{1}{p}\, \text{tr}(\mathbb{G}_{\bm{W}})\right] - \frac{2}{p}\mathbb{G}_{\bm{W}}^2 + \frac{2}{p^{*}}\mathbb{G}_{\bm{W}}\mathbb{G}_{\bm{W}^{*:}} = \bm{0}
\end{equation*}

First, we notice that since the teacher vectors are orthonormal by construction, the matrix $\mathbb{G}_{\bm{W}^{*:}}$ is an orthogonal projection matrix that projects onto the subspace spanned by the teachers. Furthermore, in order to have a solution, the matrices $\mathbb{G}_{\bm{W}}\mathbb{G}_{\bm{W}^{*:}}$ and $\mathbb{G}_{\bm{W}}^2$ must be proportional to each other $\Rightarrow$ the matrix $\mathbb{G}_{\bm{W}}$ must project onto a subspace that is itself contained within the subspace spanned by the teacher vectors ($\mathbb{G}_{\bm{W}}\mathbb{G}_{\bm{W}^{*:}} = \mathbb{G}_{\bm{W}}$).\\

Similarly to what we did before, we then have to require that

\begin{equation*}
    \mathbb{G}_{\bm{W}} = \alpha \mathbb{P}
\end{equation*}

where now the matrix $\mathbb{P}$ is an orthonormal projection onto a subspace contained within the one spanned by the teachers (i.e., all students must be linear combinations of teachers). \\

In this case the equation for $\alpha$ becomes:

\begin{equation*}
    1 - \frac{1}{p}\text{rank}(\bm{W}) - \frac{2}{p}\alpha + \frac{2}{p^{*}} = 0 \iff \alpha = \frac{p(2 + p^{*})}{p^{*}(2 + \text{rank}(\bm{W}))}
\end{equation*}

This means again that for every value of $r = \text{rank}(\bm{W}) \in [0, p^{*}]$ we have stationary points whenever the matrix $\bm{W}$ can be written as 
\begin{equation*}
    \bm{W}= \sqrt{\alpha} \mathbb{M} \mathbb{O}^{T}
\end{equation*}

where 

\begin{itemize}
    \item $\mathbb{M}$ is a generic $p \times r$ matrix such that $\mathbb{M}\mathbb{M}^T = \mathds{1}_{r}$
    \item $\mathbb{O^T}$ is a $r \times N$ matrix whose rows are orthonormal vectors ($\mathbb{O^T O} = \mathds{1}_{r}$) that are linear combination of the teacher vectors.
\end{itemize}
In terms of order parameters, these solutions correspond to:
\begin{align*}
\begin{cases}
    &\bm{Q} =\bm{W}\bm{W}^{T} = \alpha \mathbb{M} \mathbb{O}^T \mathbb{O} \mathbb{M}^T = \alpha \mathbb{M} \mathbb{M}^T\\
    &\bm{\rho} \bm{\rho}^T = \bm{W}\bm{W}\bm{W}^{*T} \bm{W}^{*} \bm{W}^{T} = \alpha \mathbb{M} \mathbb{O}^{T} \bm{W}\bm{W}^{*T} \bm{W}^{*} \mathbb{O} \mathbb{M}^{T} = \alpha \mathbb{M} \mathbb{M}^{T} = \bm{Q}
\end{cases}
\end{align*}

\paragraph{Example 3: zero error solution ($r = \text{rank}(W) = p^{*}$)\\}

In this case $\alpha = p/p^{*}$ and the matrix $\mathbb{M}$ corresponds to a generic $p \times p^{*}$ rectangular rotation. In this case we have:

\begin{equation*}
    \bm{W}= \sqrt{\frac{p}{p^{*}}} \mathbb{M} \bm{W}^{*:}
\end{equation*}

and

\begin{equation*}
    \mathbb{G}_{W} = \bm{W}^T \bm{W}= \frac{p}{p^{*}} \bm{W}^{*T} \bm{W}^{*:} = \frac{p}{p^{*}} \mathbb{G}_{\bm{W}^{*:}}
\end{equation*}

which means that the student is completely equivalent to the teacher network. We also note that this solution corresponds to the $\text{tr}(\mathbb{G}_{W}) = p$ solution.

\paragraph{Example 4: all students aligned to one teacher ($r = \text{rank}(W) = 1$)\\}

In this case $\alpha = \frac{p(2+p^{*})}{3 p^{*}}$, the matrix $\mathbb{M}$ corresponds to a generic $p$-dimensional vector $\bm{v}$ with norm one and the overlap matrix is equal to:

\begin{equation*}
    \bm{Q} = \alpha \mathbb{M} \mathbb{M}^T = \alpha \bm{v} \bm{v}^{T}
\end{equation*}

which corresponds to students all aligned to the same teacher with quadratic norm $Q_{ii} = \alpha v_{i}^2$ such that $\sum_{k=1}^{p} Q_{ii} = \alpha \sum_{k=1}^{p} v_{i}^2 = \alpha = \frac{p(2+p^{*})}{3 p^{*}}$. 

\subsubsection{Study of the Hessian}

We can now study the Hessian matrix and its eigenvalues to understand the nature of the stationary points found above. We notice that for this model the Hessian is an $p N \times p N$ matrix, structured as an $p \times p$ block matrix, where each block corresponds to a pair of students and is of size $N \times N$. Therefore, it will also be useful to look for $p N$-dimensional eigenvectors in the form of $p$ vectors, each of dimension $N$, corresponding to the different student perceptrons. We will study explicitly all the examples presented above, keeping in mind that they don't include all the stationary points but include the most relevant points for dynamics.

\paragraph{Tabula rasa initialization}

For this stationary points all the students weights are equal to zero (and in particular $\bm{Q} = \bm{0}$, $\bm{\rho} = \bm{0}$) and the Hessian matrix simplifies to

\begin{equation*}
    \nabla_{\vec{w}_p} \nabla_{\vec{w}_q} \mathcal{R}(\{\mathcal{Q}\}) = \mathcal{H}_{pq} = -\frac{4}{p}\delta_{qp}\left[\mathds{1}_{N} + \frac{2}{p^{*}}\sum_{l=1}^{p^{*}}\vec{w}^{*}_{l}\vec{w}^{*T}_{l}\right]
\end{equation*}

One can show that the eigenvalues and eigenvectors of this matrix are structured like this:\\

\paragraph{Eigenvalues distribution for tablua rasa initialization\\}

\underline{$pN$ NEGATIVE EIGENVALUES}

\[
\left\{
\begin{aligned}
&\bullet \; -\frac{4}{p} \left[1 + \frac{2}{p^{*}} \right]; \ \text{multiplicity $pp^{*}$, associated to } 
\begin{pmatrix}
\vec{w}^{*}_{l} \\
\vec{0} \\
\vdots \\
\vec{0}
\end{pmatrix}
, \dots,
\begin{pmatrix}
\vec{0} \\
\vdots \\
\vec{0} \\
\vec{w}^{*}_{l}
\end{pmatrix}, \quad \forall l \in [1, p^{*}] \\
&\bullet \; -\frac{4}{p}; \ \text{multiplicity $p(N-p^{*})$, associated to } 
\begin{pmatrix}
\vec{v} \\
\vec{0} \\
\vdots \\
\vec{0}
\end{pmatrix}
, \dots,
\begin{pmatrix}
\vec{0} \\
\vdots \\
\vec{0} \\
\vec{v}
\end{pmatrix}
, \quad \vec{v} \perp \vec{w}^{*}_l \quad \forall l \in [1,p^{*}]
\end{aligned}
\right.
\]

As one might expect, this points corresponds to a local maximum.

\paragraph{Plateau region}

These stationary points correspond to the case where all the students are orthogonal to each other and to the teacher. Their norm is fixed at $p/(2 + p)$ by the stationary condition. This region corresponds to the \textit{plateau} where the dynamics of online learning initially get stuck before starting to recover the signal. Here $\bm{\rho} = \bm{0}$ and the matrix $\bm{Q}$ is a diagonal matrix with all entries equal to $p/(2 + p)$.\\

The Hessian matrix is equal to:
\begin{equation*}
    \mathcal{H}_{pq} = -\frac{8}{p}\left\{-\frac{1}{p}\left(\vec{w}_{q}\vec{w}_{p}^{T} + \vec{w}_{p}\vec{w}_{q}^{T}\right) + \delta_{qp} \left[ \frac{1}{p^{*}} \sum_{l=1}^{p^{*}} \vec{w}^{*}_{l} \vec{w}^{*T}_{l} - \frac{1}{p} \vec{w}_{k}\vec{w}_{k}^{T}\right]\right\} 
\end{equation*}

In this case the eigenvalues and eigenvectors of this matrix are:\\

\paragraph{Eigenvalues distribution for the plateau region\\}

\underline{$pp^{*}$ NEGATIVE EIGENVALUES}

\[
\left\{
\begin{aligned}
&\bullet \; -\frac{8}{pp^{*}}; \ \text{multiplicity $pp^{*}$, associated to } 
\begin{pmatrix}
\vec{w}^{*}_{l} \\
\vec{0} \\
\vdots \\
\vec{0}
\end{pmatrix}
, \dots,
\begin{pmatrix}
\vec{0} \\
\vdots \\
\vec{0} \\
\vec{w}^{*}_{l}
\end{pmatrix}, \quad \forall l \in [1, p^{*}] 
\end{aligned}
\right.
\]

\underline{$p(p+1)/2$ POSITIVE EIGENVALUES}

\[
\left\{
\begin{aligned}
&\bullet \; \frac{8}{p}; \ \text{multiplicity $1$, associated to } 
\begin{pmatrix}
\vec{w}_{1} \\
\vec{w}_{2} \\
\vdots \\
\vec{w}_{p}
\end{pmatrix}
 \\
&\bullet \; \frac{16}{p(p+2)}; \ \text{multiplicity $\frac{p(p-1)}{2} + p - 1$, associated to } 
\begin{pmatrix}
\vec{0} \\
\vdots \\
\vec{w}_{k} \quad \text{(index $k'$)}\\
\vdots \\
\vec{w}_{k'} \quad \text{(index $k$)}\\
\vdots \\
\vec{0}
\end{pmatrix}
,
\begin{pmatrix}
\vec{0} \\
\vdots \\
\vec{w}_{k} \quad \text{(index $k$)}\\
\vdots \\
-\vec{w}_{k'} \quad \text{(index $k'$)}\\
\vdots \\
\vec{0}
\end{pmatrix}
\end{aligned}
\right.
\]

\underline{$p(N-p^{*}-p) + p(p-1)/2 $ NULL EIGENVALUES}

\[
\left\{
\begin{aligned}
&\bullet \; \text{multiplicity $p(N-p^{*}-p)$, associated to } 
\begin{pmatrix}
\vec{v} \\
\vec{0} \\
\vdots \\
\vec{0}
\end{pmatrix}
, \dots,
\begin{pmatrix}
\vec{0} \\
\vdots \\
\vec{0} \\
\vec{v}
\end{pmatrix}
, \quad \vec{v} \perp \vec{w}^{*}_l \perp \vec{w}_k \quad \forall l \in [1,p^{*}], \, \forall k \in [1,p]
 \\
&\bullet \; \text{multiplicity $\frac{p(p-1)}{2}$, associated to } 
\begin{pmatrix}
\vec{0} \\
\vdots \\
\vec{w}_{k} \quad \text{(index $k'$)}\\
\vdots \\
-\vec{w}_{k'} \quad \text{(index $k$)}\\
\vdots \\
\vec{0}
\end{pmatrix}
\end{aligned}
\right.
\]

We notice that the ratio between negative and null eigenvalues is equal to:

\begin{equation*}
    \frac{\textit{\# NEGATIVE EIGENVALUES}}{\textit{\# NULL EIGENVALUES}} = \frac{p^{*}}{N - p^{*} - p/2 - 1/2}
\end{equation*}

that grows with $p$ up to $p = N$ where
\begin{equation*}
    \frac{\textit{\# NEGATIVE EIGENVALUES}}{\textit{\# NULL EIGENVALUES}} = \frac{p^{*}}{N/2 - p^{*} - 1/2}
\end{equation*}

\paragraph{Zero error solution}

These stationary points correspond to a student network completely equivalent to the teacher and to the final points of the online learning dynamics (in the noiseless case). All the students are linear combinations of teacher perceptrons and the student matrix can be written as

\begin{equation*}
    \bm{W}= \sqrt{\frac{p}{p^{*}}} \mathbb{S} \bm{W}^{*:}
\end{equation*}

where $\mathbb{S}$ is a generic $p \times p^{*}$ matrix such that

\begin{equation*}
    \mathbb{S}^{T}\mathbb{S} = \mathds{1}_{p^{*}}.
\end{equation*}

The Hessian matrix is equal to:
\begin{equation*}
    \mathcal{H}_{pq} = \frac{8}{p p^{*}} \left\{\sum_{l} \mathbb{S}_{ql} \mathbb{S}_{pl} \mathds{1}_{N} + \sum_{l, l'} \mathbb{S}_{ql} \mathbb{S}_{pl'} \left( \vec{w}^{*}_{l}\vec{w}^{*T}_{l'} + \vec{w}^{*}_{l'}\vec{w}^{*T}_{l} \right) \right\} 
\end{equation*}

In this case the eigenvalues and eigenvectors of this matrix are:\\

\textbf{Eigenvalues distribution for the solution\\}

\underline{$p^{*}(N-p^{*}) + p^{*}(p^{*}+1)/2$ POSITIVE EIGENVALUES}

\[
\left\{
\begin{aligned}
&\bullet \; \frac{8}{pp^{*}}; \ \text{multiplicity $p^{*}(N-p^{*})$, associated to } 
\begin{pmatrix}
x_1 \vec{v} \\
x_2 \vec{v} \\
\vdots \\
x_R \vec{v} \\
\end{pmatrix}\\
&\text{with} \quad \vec{v} \perp \vec{w}^{*}_l \quad \forall l \in [1,p^{*}] \quad \text{and} \quad \vec{x} \in \{\text{\footnotesize{positive eigenvalues of} } \mathbb{S} \mathbb{S}^{T} \}
 \\
&\bullet \; \frac{16}{p p^{*}}; \ \text{multiplicity $p^{*}(p^{*}+1)/2 - 1$, associated to} 
\begin{pmatrix}
\sum_{j=1}^{p^{*}} \mathbb{X}_{1j} \vec{w}^{*}_j \\
\sum_{j=1}^{p^{*}} \mathbb{X}_{2j} \vec{w}^{*}_j \\
\vdots \\
\sum_{j=1}^{p^{*}} \mathbb{X}_{Rj} \vec{w}^{*}_j
\end{pmatrix},\\
&\text{with} \quad \mathbb{X} = \mathbb{S} \mathbb{P}, \quad \mathbb{P}^{T} = \mathbb{P}, \quad \text{tr}(\mathbb{P}) = 0
 \\
&\bullet \; \frac{8}{p p^{*}} (2 + p^{*}); \ \text{multiplicity 1, associated to} 
\begin{pmatrix}
\sum_{j=1}^{p^{*}} \mathbb{S}_{1j} \vec{w}^{*}_j \\
\sum_{j=1}^{p^{*}} \mathbb{S}_{2j} \vec{w}^{*}_j \\
\vdots \\
\sum_{j=1}^{p^{*}} \mathbb{S}_{Rj} \vec{w}^{*}_j
\end{pmatrix} = \; \begin{pmatrix}
\vec{w}_{1} \\
\vec{w}_{2} \\
\vdots \\
\vec{w}_{p} \\
\end{pmatrix}\\ 
\end{aligned}
\right.
\]

\underline{$(N-p^{*})(p-p^{*}) + p^{*}(p-p^{*}) + p^{*}(p^{*}-1)/2$ NULL EIGENVALUES}

\[
\left\{
\begin{aligned}
&\bullet \; \text{multiplicity $(N-p^{*})(p-p^{*})$, associated to } 
\begin{pmatrix}
x_1 \vec{v} \\
x_2 \vec{v} \\
\vdots \\
x_R \vec{v} \\
\end{pmatrix}, \\
&\text{with} \quad \vec{v} \perp \vec{w}^{*}_l \quad \forall l \in [1,p^{*}] \quad \text{and} \quad \vec{x} \in \{\text{\footnotesize{null eigenvalues of} } \mathbb{S} \mathbb{S}^{T} \}
 \\
&\bullet \; \text{multiplicity $p^{*}(p-p^{*})$, associated to} 
\begin{pmatrix}
x_1 \vec{w}^{*}_{l} \\
x_2 \vec{w}^{*}_{l} \\
\vdots \\
x_R \vec{w}^{*}_{l} \\
\end{pmatrix}, \\
&\forall l \in [1,p^{*}] \quad \text{and} \quad \vec{x} \in \{\text{\footnotesize{null eigenvalues of} } \mathbb{S} \mathbb{S}^{T} \}
 \\
&\bullet \; \text{multiplicity $p^{*}(p^{*}-1)/2$, associated to} 
\begin{pmatrix}
\sum_{j=1}^{p^{*}} \mathbb{X}_{1j} \vec{w}^{*}_j \\
\sum_{j=1}^{p^{*}} \mathbb{X}_{2j} \vec{w}^{*}_j \\
\vdots \\
\sum_{j=1}^{p^{*}} \mathbb{X}_{Rj} \vec{w}^{*}_j
\end{pmatrix},\\
&\text{with} \quad \mathbb{X} = \mathbb{S} \mathbb{P}, \quad \mathbb{P}^{T} = - \mathbb{P}
\end{aligned}
\right.
\]

As we expected, these points don't have negative directions since they correspond to the global minima of the loss landscape. 

\subsubsection{Tangent flat directions at solutions}
\label{sec:tangent_flat}

We now want to distinguish, among the eigenvectors found above, those associated to true flat directions of the manifold of zero error solutions from those that are only Hessian-flat. The manifold of exact zero error solutions is

\begin{equation*}
    \bm{W}= \sqrt{\frac{p}{p^{*}}} \mathbb{S} \bm{W}^{*:}, \qquad \mathbb{S}^{T}\mathbb{S} = \mathds{1}_{p^{*}}.
\end{equation*}

If we perturb the matrix $\mathbb{S}$ by $\delta \mathbb{S}$, in order to remain on the manifold at first order we must impose

\begin{equation*}
    \delta(\mathbb{S}^{T}\mathbb{S}) = \mathbb{S}^{T}\delta\mathbb{S} + \delta\mathbb{S}^{T}\mathbb{S} = \bm{0}.
\end{equation*}

The general solution of this constraint can be written as

\begin{equation*}
    \delta\mathbb{S} = \mathbb{S}\Omega + \mathbb{S}_{\perp}K
\end{equation*}

where

\begin{itemize}
    \item $\Omega$ is a $p^{*}\times p^{*}$ antisymmetric matrix, $\Omega^{T} = - \Omega$
    \item $\mathbb{S}_{\perp}$ is a $p \times (p-p^{*})$ matrix whose columns form an orthonormal basis of the subspace orthogonal to the columns of $\mathbb{S}$
    \item $K$ is a generic $(p-p^{*})\times p^{*}$ matrix.
\end{itemize}

The corresponding perturbation of the student matrix is

\begin{equation*}
    \delta\bm{W} = \sqrt{\frac{p}{p^{*}}} \delta\mathbb{S} \bm{W}^{*:} = \sqrt{\frac{p}{p^{*}}}\left(\mathbb{S}\Omega + \mathbb{S}_{\perp}K\right)\bm{W}^{*:}.
\end{equation*}

We can now identify the two families of eigenvectors with null eigenvalues.

\paragraph{First tangent family: directions generated by $\mathbb{S}_{\perp}K$\\}

For fixed $l \in [1,p^{*}]$, by taking the $l$-th column of $K$ equal to a generic vector and all the others equal to zero, we obtain perturbations of the form

\begin{equation*}
    \delta\bm{W} =
    \begin{pmatrix}
        x_1 \vec{w}^{*}_{l} \\
        x_2 \vec{w}^{*}_{l} \\
        \vdots \\
        x_R \vec{w}^{*}_{l}
    \end{pmatrix}
\end{equation*}

with $\vec{x}$ in the column space of $\mathbb{S}_{\perp}$, which is exactly the null eigenspace of $\mathbb{S}\mathbb{S}^{T}$. Therefore these perturbations coincide with the second null family found in the Hessian analysis. Their multiplicity is

\begin{equation*}
    p^{*}(p-p^{*}).
\end{equation*}

\paragraph{Second tangent family: directions generated by $\mathbb{S}\Omega$\\}

If we instead perturb with

\begin{equation*}
    \delta\mathbb{S} = \mathbb{S}\Omega, \qquad \Omega^{T} = -\Omega,
\end{equation*}

we obtain

\begin{equation*}
    \delta\bm{W} = \sqrt{\frac{p}{p^{*}}}\mathbb{S}\Omega \bm{W}^{*:}.
\end{equation*}

Defining $\mathbb{X} = \mathbb{S}\Omega$, this can be rewritten as

\begin{equation*}
    \delta\bm{W} =
    \begin{pmatrix}
        \sum_{j=1}^{p^{*}} \mathbb{X}_{1j} \vec{w}^{*}_{j} \\
        \sum_{j=1}^{p^{*}} \mathbb{X}_{2j} \vec{w}^{*}_{j} \\
        \vdots \\
        \sum_{j=1}^{p^{*}} \mathbb{X}_{Rj} \vec{w}^{*}_{j}
    \end{pmatrix},
    \qquad
    \mathbb{X} = \mathbb{S}\Omega, \qquad \Omega^{T} = -\Omega,
\end{equation*}

which is exactly the third null family found above. Its multiplicity is

\begin{equation*}
    \frac{p^{*}(p^{*}-1)}{2}.
\end{equation*}

Summing the dimensions of these two tangent families, we obtain

\begin{equation*}
    p^{*}(p-p^{*}) + \frac{p^{*}(p^{*}-1)}{2}
    =
    pp^{*} - \frac{p^{*}(p^{*}+1)}{2},
\end{equation*}

which is exactly the dimension of the solution manifold.

\paragraph{Null directions that are not tangent to the manifold\\}

The remaining null family is given by perturbations of the form

\begin{equation*}
    \delta\bm{W} =
    \begin{pmatrix}
        x_1 \vec{v} \\
        x_2 \vec{v} \\
        \vdots \\
        x_R \vec{v}
    \end{pmatrix},
    \qquad
    \vec{v} \perp \vec{w}^{*}_{l} \quad \forall l \in [1,p^{*}],
    \qquad
    \vec{x} \in \{\text{null eigenvalues of } \mathbb{S}\mathbb{S}^{T}\}.
\end{equation*}

These perturbations are Hessian-flat, but they are not generated by a perturbation $\delta \mathbb{S}$ and therefore are not connected to the continuous symmetry of the problem.
\paragraph{All the students aligned to one teacher}

We now study the case where all the students are aligned to one of teacher (for example, $\vec{w}^{*}_{1}$). Once again, this point is not relevant in the random initialization dynamics. The sum of the norms of the students is fixed to $\frac{p(2+p^{*})}{3 p^{*}}$ by the stationarity condition.\\

The Hessian matrix is equal to:
\begin{equation*}
    \mathcal{H}_{pq} = -\frac{8}{p}\left\{-\frac{1}{p} \vec{Q}_{q} \vec{Q}_{p}\left(\mathds{1}_{N} + 2 \vec{w}^{*}_{1}\vec{w}^{*T}_{1}\right) + \delta_{qp} \left[ \frac{p^{*} - 1}{3 p^{*}} \mathds{1}_{N} + \frac{1}{p^{*}}\sum_{l=1}^{p^{*}}\vec{w}^{*}_{l}\vec{w}^{*T}_{l} -\frac{2 + p^{*}}{3 p^{*}}\vec{w}^{*}_{1}\vec{w}^{*T}_{1}\right]\right\} 
\end{equation*}

where we defined the vector $\vec{Q} = \begin{pmatrix}
\sqrt{Q_{11}} \\
\sqrt{Q_{22}} \\
\vdots \\
\sqrt{Q_{RR}}
\end{pmatrix} \in \mathbb{R}^{p}$\\

\textbf{Eigenvalues distribution for the all aligned to $\vec{w}^{*}_{1}$ region ($p^{*} > 1$)\\}

\underline{$(N-1)(p-1)$ NEGATIVE EIGENVALUES}

\[
\left\{
\begin{aligned}
&\bullet \; -\frac{8 (p^{*} - 1)}{3 p p^{*}}; \ \text{multiplicity $(N -p^{*})(p -1)$, associated to} 
\begin{pmatrix}
x_{1} \vec{v} \\
x_{2} \vec{v}  \\
\vdots \\
x_{p} \vec{v} 
\end{pmatrix}
\text{with} \; \vec{x} \cdot \vec{q} = 0, \, \vec{v} \perp \hat{j}, \vec{w}^{*}_{l} \; \forall l \in [1, p^{*}] \\
&\bullet \; -\frac{8 (p^{*} + 2)}{3 p p^{*}}; \ \text{multiplicity $(p^{*}-1)(p-1)$, associated to} 
\begin{pmatrix}
x_{1} \vec{T_{l}} \\
x_{2} \vec{T_{l}}  \\
\vdots \\
x_{p} \vec{T_{l}} 
\end{pmatrix}
\text{with} \; \vec{x} \cdot \vec{q} = 0, \, \; \forall l \in [2, p^{*}] \\
\end{aligned}
\right.
\]

\underline{$N - p^{*} + 1$ POSITIVE EIGENVALUE}

\[
\left\{
\begin{aligned}
&\bullet \; \frac{8}{p p^{*}}; \ \text{multiplicity $(N-p^{*})$, associated to } 
\begin{pmatrix}
q_{1} \vec{v} \\
q_{2} \vec{v}\\
\vdots \\
q_{p} \vec{v}
\end{pmatrix}
\text{with} \; \vec{v} \perp \vec{w}^{*}_{l}, \vec{w}^{*}_{l} \; \forall l \in [1, p^{*}] \\
&\bullet \; \frac{8 (2 + p^{*})}{p p^{*}}; \ \text{multiplicity $1$, associated to } 
\begin{pmatrix}
q_{1} \vec{w}^{*}_{1} \\
q_{2} \vec{w}^{*}_{1}\\
\vdots \\
q_{p} \vec{w}^{*}_{1}
\end{pmatrix} = 
\begin{pmatrix}
\vec{w}_{1} \\
\vec{w}_{2}\\
\vdots \\
\vec{w}_{p}
\end{pmatrix} 
\end{aligned}
\right.
\]

\underline{$p + p^{*} - 2$ NULL EIGENVALUES}

\[
\left\{
\begin{aligned}
&\bullet \; \text{multiplicity $p - 1$, associated to } 
\begin{pmatrix}
x_{1} \vec{w}^{*}_{1} \\
x_{2} \vec{w}^{*}_{1}\\
\vdots \\
x_{p} \vec{w}^{*}_{1}
\end{pmatrix}
\text{with} \; \vec{x} \cdot \vec{q} = 0\\
&\bullet \; \text{multiplicity $p^{*} - 1$, associated to } 
\begin{pmatrix}
q_{1} \vec{T_{l}} \\
q_{2} \vec{T_{l}}  \\
\vdots \\
q_{p} \vec{T_{l}} 
\end{pmatrix}
\text{with} \; \forall l \in [2, p^{*}] \\
\end{aligned}
\right.
\]

For $p = p^{*} = 1$ this region is equivalent to the \textit{zero error solution} and one can check that for these values of $p$ and $p^{*}$, the two results for eigenvalues and eigenvectors coincide.

\subsection{Time of escape}

In this section, we show how to obtain the 
average time of escape from the plateau region shown in the main text. The dynamic equations in the plateau region are:
\begin{equation}
    \begin{cases}
      \displaystyle\frac{d\rho_{kl}}{dt} = \omega_{\rho}(p,p^{*}) \rho_{kl} + \mathcal{O}(\epsilon^{2}) + \mathcal{O}(\eta^2)  \quad \forall k \in [1, p], \forall l \in [1, p^{*}] \\[1em]
       \displaystyle \frac{dQ_{kk'}}{dt} = \omega_{Q}(p)Q_{kk'} + \mathcal{O}(\epsilon^{2}) + \mathcal{O}(\eta^2) \quad \forall k, k' \in [1, p], k \neq k' \\[1em]
       \displaystyle \frac{dQ_{kk}}{dt} = \mathcal{O}(\epsilon^2) + \mathcal{O}(\eta^{2}); \quad Q_{kk} = \Bar{Q} + \mathcal{O}(\epsilon^{2}) + \mathcal{O}(\eta)\quad \forall k \in [1, p]\\
      \end{cases}\,.
      \label{eq:plateau_odes2}
\end{equation}

In this region, the population risk is equal to:

\begin{align*}
\mathcal{R}(\{\mathcal{Q}\}) &= \mathbb{E}_{x} \left[ \ell(\{u_{l}\}, \{\lambda_{k}\}, \vec{x}) \right] = \\
 &= \left[1+ \frac{2}{p^{*}} - \frac{p}{p+2}\right] + \frac{2}{p^2}\exp{(-2 \omega_{Q}t)}\sum_{r_1 \neq r_2}Q_{r_1 r_2}^2(0) - \frac{4}{p p^*}\exp{(2 \omega_{\rho}t})\sum_{r \sigma}\rho_{r\sigma}^2(0),
\end{align*}

where 

\begin{equation}
    \omega_{\rho}(p, p^{*}) = \frac{8}{p p^{*}}, \quad \omega_{Q}(p) = \frac{16}{p(p+2)}.
\end{equation}

We call $\mu_{0} = N \sum_{r, \sigma} \rho_{r \sigma}^2 (0)$,which is $\mathcal{O}(1)$ if $\rho(0), Q(0) \sim 1/\sqrt{N}$, and $\tau_{0} = N \sum_{r_1 \neq r_2} Q_{r_1, r_2}^2 (0)$, which is zero with our orthogonal initialization. 
If we start the dynamics at the plateau, we can write the evolution of the population risk:

\begin{equation*}
    \mathcal{R}(t) = \mathcal{R}(0) - \frac{1}{N}\frac{2\tau_0}{p^2}[1 - \exp{(-2\omega_{Q}t)}] - \frac{1}{N}\frac{4 \mu_{0}}{p p^{*}}[\exp{(2\omega_{\rho}t) - 1}],
\end{equation*}

where 

\begin{equation*}
    \mathcal{R}(0) = 1+ \frac{2}{p^{*}} - \frac{p}{p+2} + \frac{1}{N}\left(\frac{2\tau_0}{p^2} - \frac{4 \mu_{0}}{pp^{*}}\right).
\end{equation*}

The relative difference, remembering that $\tau_0=0$ for the orthogonal initialization used in the main text, is then given by:
\begin{equation*}
    \frac{\mathcal{R}(0) - \mathcal{R}(t)}{\mathcal{R}(0)} 
    = 
    \frac{\left( - \frac{4\mu_0}{p p^*}\right) + \frac{4\mu_0}{p p^*}\exp\!\left(\frac{16}{p p^{*}}t\right)}
         {N\!\left(1+ \frac{2}{p^*} - \frac{p}{p+2}\right) - \frac{4\mu_0}{p p^*}} .
\end{equation*}

Assuming that the $\rho_{r\sigma}(0)$ are Gaussian variables $\mathcal{N}(0, 1/N)$ (while $Q_{r_1 r_2}(0) = 0$), as we do in the main text, we have $\langle\mu_0\rangle = p p^{*}$. By substituting these values into the formula and rescaling time as $\tilde t = t/p$ (as expained in the main text), we obtain
\begin{equation*}
    \left\langle 
    \frac{\mathcal{R}(0) - \mathcal{R}(t)}{\mathcal{R}(0)}
    \right\rangle 
    = 
    \frac{
        4 \left( e^{\frac{16}{p^{*}} \tilde t} - 1 \right)
    }{
        N \left( 1 + \frac{2}{p^{*}} - \frac{p}{p + 2} \right) 
    } + \mathcal{O}\!\left(\frac{1}{N^2}\right).
\end{equation*}

\subsection{Conservation of $S(t)$}
\label{app:S_conservation}

In this appendix, we provide an argument showing that the matrix $S(t)$, defined as
\begin{equation}
    S(t) = \boldsymbol{\rho}(t) \left[ \boldsymbol{\rho}(t)^T \boldsymbol{\rho}(t) \right]^{-1/2},
\end{equation}
is conserved during the learning dynamics. In the continuous-time limit ($\eta \ll 1$), the ordinary differential equations governing the overlaps can be written in the form
\begin{align}
    \frac{d\boldsymbol{\rho}}{dt} &= c(t) \boldsymbol{\rho} - \frac{8}{p^2} \mathbf{Q} \boldsymbol{\rho}, \label{eq:rho_exact_ode} \\
    \frac{d\mathbf{Q}}{dt} &= a(t) \mathbf{Q} + \frac{16}{p p^*} \boldsymbol{\rho} \boldsymbol{\rho}^T - \frac{16}{p^2} \mathbf{Q}^2. \label{eq:Q_exact_ode}
\end{align}
where $a(t)$ and $c(t)$ are scalar functions. We can rewrite Eq.~\eqref{eq:rho_exact_ode} as:
\begin{equation}
    \frac{d\boldsymbol{\rho}}{dt} = H(t) \boldsymbol{\rho}(t), \quad \text{where} \quad H(t) \equiv c(t)\mathbf{I}_p - \frac{8}{p^2} \mathbf{Q}(t).
\end{equation}

Under orthogonal initialization, $\mathbf{Q}(0) = \mathbf{I}_p$. The structure of Eq.~\eqref{eq:Q_exact_ode} suggests the folllwing property: since $\mathbf{Q}$ starts as the identity matrix, the discrete updates generated by the coupled equations only involve matrix products of $\mathbf{Q}$ and $\boldsymbol{\rho}\boldsymbol{\rho}^T$, and therefore do not introduce new eigendirections beyond those generated by the initial matrix $\boldsymbol{\rho}(0)\boldsymbol{\rho}(0)^T$. We therefore use the following assumption, supported by the numerical conservation shown below: for all $t>0$, both $\mathbf{Q}(t)$ and $H(t)$ remain symmetric matrix functions of $\boldsymbol{\rho}(0)\boldsymbol{\rho}(0)^T$. Under this assumption, the time evolution operator is simply a matrix function $f$ evaluated on this initial state, and the solution for $\boldsymbol{\rho}(t)$ can be written as
\begin{equation}
    \boldsymbol{\rho}(t) = f\left(\boldsymbol{\rho}_0\boldsymbol{\rho}_0^T\right) \boldsymbol{\rho}_0,
\end{equation}
where $\boldsymbol{\rho}_0 \equiv \boldsymbol{\rho}(0)$.

For any analytic function $f$ that admits a power-series expansion, and any
rectangular matrix $A \in \mathbb{R}^{m \times n}$, the identity $f(AA^T)A = Af(A^TA)$ holds.
Here the same function $f$ is applied as a matrix function to two different
symmetric matrices, $AA^T \in \mathbb{R}^{m\times m}$ and
$A^TA \in \mathbb{R}^{n\times n}$, so the output spaces of the two matrix operators are generally different. This allows us to pull $\boldsymbol{\rho}_0$ to the front:
\begin{equation}
    \boldsymbol{\rho}(t) = \boldsymbol{\rho}_0 f\left(\boldsymbol{\rho}_0^T \boldsymbol{\rho}_0\right).
\end{equation}
Let $D(t) \equiv f\left(\boldsymbol{\rho}_0^T \boldsymbol{\rho}_0\right)$. Because $D(t)$ is simply a polynomial of $\boldsymbol{\rho}_0^T \boldsymbol{\rho}_0$, it is a symmetric matrix that commutes with $\boldsymbol{\rho}_0^T \boldsymbol{\rho}_0$. The evolution of the overlap matrix is $\boldsymbol{\rho}(t) = \boldsymbol{\rho}_0 D(t)$.

Substituting this directly into the definition of $S(t)$:
\begin{align}
    S(t) &= \left( \boldsymbol{\rho}_0 D \right) \left[ \left(\boldsymbol{\rho}_0 D\right)^T \left(\boldsymbol{\rho}_0 D\right) \right]^{-1/2} \\
         &= \boldsymbol{\rho}_0 D \left[ D \boldsymbol{\rho}_0^T \boldsymbol{\rho}_0 D \right]^{-1/2}.
\end{align}
Because $D$ and $\boldsymbol{\rho}_0^T \boldsymbol{\rho}_0$ commute, we can rearrange the terms inside the bracket to $D^2 \boldsymbol{\rho}_0^T \boldsymbol{\rho}_0$. The inverse square root distributes cleanly over these commuting positive-definite matrices:
\begin{align}
    S(t) &= \boldsymbol{\rho}_0 D \left[ D^2 \right]^{-1/2} \left[ \boldsymbol{\rho}_0^T \boldsymbol{\rho}_0 \right]^{-1/2} \\
         &= \boldsymbol{\rho}_0 D D^{-1} \left[ \boldsymbol{\rho}_0^T \boldsymbol{\rho}_0 \right]^{-1/2} \\
         &= \boldsymbol{\rho}_0 \left[ \boldsymbol{\rho}_0^T \boldsymbol{\rho}_0 \right]^{-1/2} \\
         &= S(0).
\end{align}
Under the closure assumption above, this shows that $S(t) = S(0)$ for all time $t \geq 0$, implying that the trajectory remains on the manifold dictated by the random initialization.

\begin{figure}[h]
    \centering
    \includegraphics[width=0.6\linewidth]{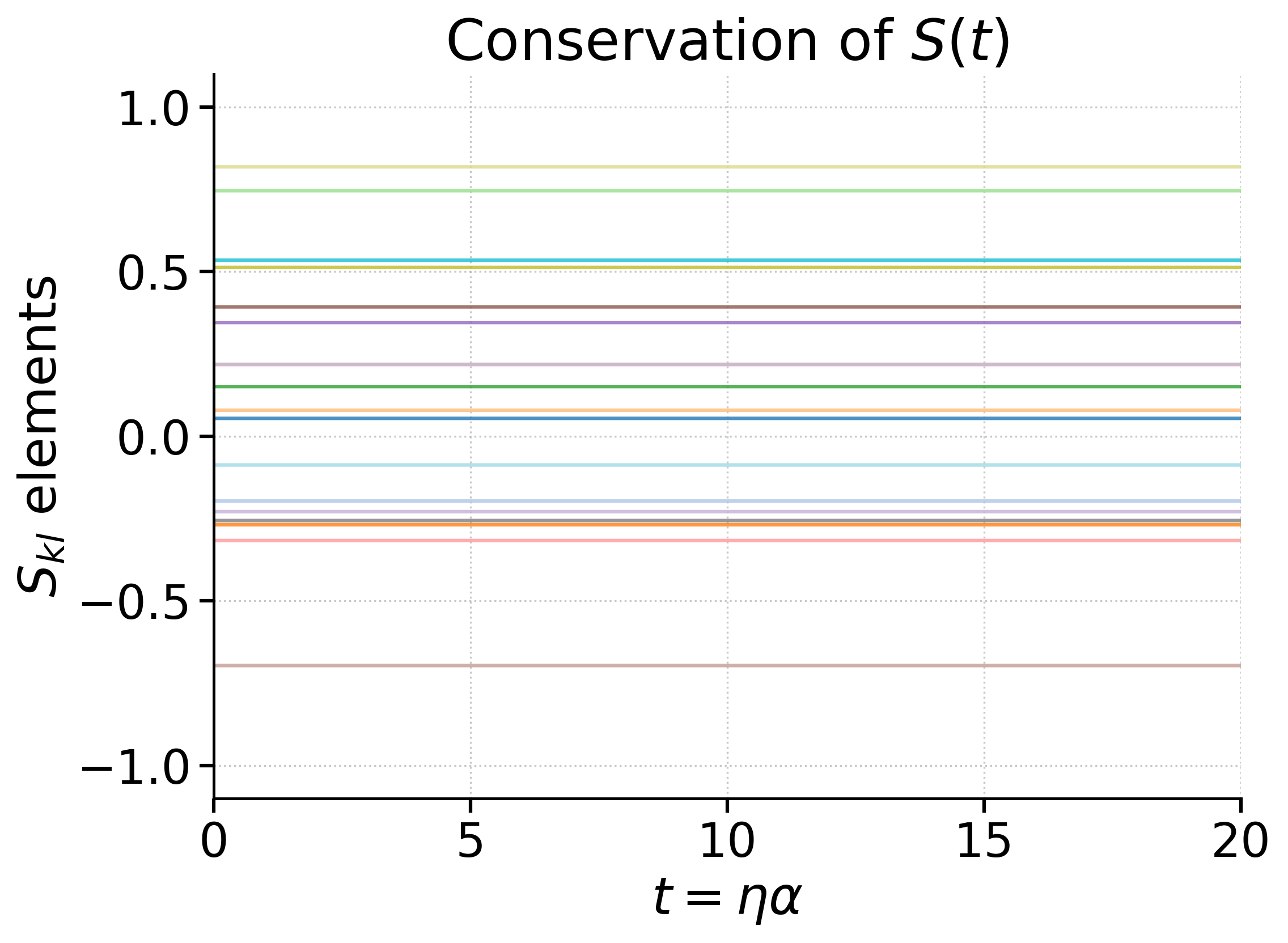}
    \caption{Evolution of the elements of the matrix $S(t)$ over time for a random orthogonal initialization, illustrating their numerical conservation during the learning dynamics.}
    \label{fig:S_matrix_conservation}
\end{figure}

In Figure~\ref{fig:S_matrix_conservation}, we show for a random initialization that the elements of $S(t)$ remain constant during learning.

\subsection{Study of the degrees of freedom}

\subsubsection{Student network expressivity}

Since we know that, for networks with quadratic activation, the outputs are fully determined by the matrix $G_{W} = W^{T}W$, which is a \(N \times N\) matrix, we define the student's \textit{degrees of freedom} as the dimension of the subspace of $G_{W}$ that the student can span, which is also an indicator for the network expressivity. By definition, the student's degrees of freedom is a number $\textit{DOF} \in [0, (N(N+1)/2]$, given that $G_{W}$ is symmetric. We will study how this quantity varies as a function of the network parameters. 

\paragraph{Case $p \leq N$.} This includes the scenario that we studied in the previous sections. In this case, we have
\begin{itemize}
    \item $N p$ initial degrees of freedom to construct the matrix $W$,
    \item $p(p-1)/2$ rotations that leave the final $G_{W}$ unchanged.
\end{itemize}

The effective number of degrees of freedom is then given by
\begin{equation*}
    \textit{DOF} = Np - p(p-1)/2, \quad \text{if} \quad p \leq N.
\end{equation*}

\paragraph{Case $p > N$.} This corresponds to the scenario of a student network where the hidden layer dimension is bigger than the input dimension. In this case, we have

\begin{itemize}
    \item $N p$ initial degrees of freedom to construct the matrix $W$,
    \item $p(p-1)/2$ rotations that leave the final $G_{W}$ unchanged.
    \item Since $p > N$, only $N$ of the columns of $\mathbb{W}$ are independent $\Rightarrow  (p-N)(p-N-1)/2$ rotations act on the subspace formed by the remaining $(p-N)$ vectors, leaving $\mathbb{W}$ itself unchanged.
\end{itemize}

\begin{figure}[h!]
    \centering
    \includegraphics[width=0.6\linewidth]{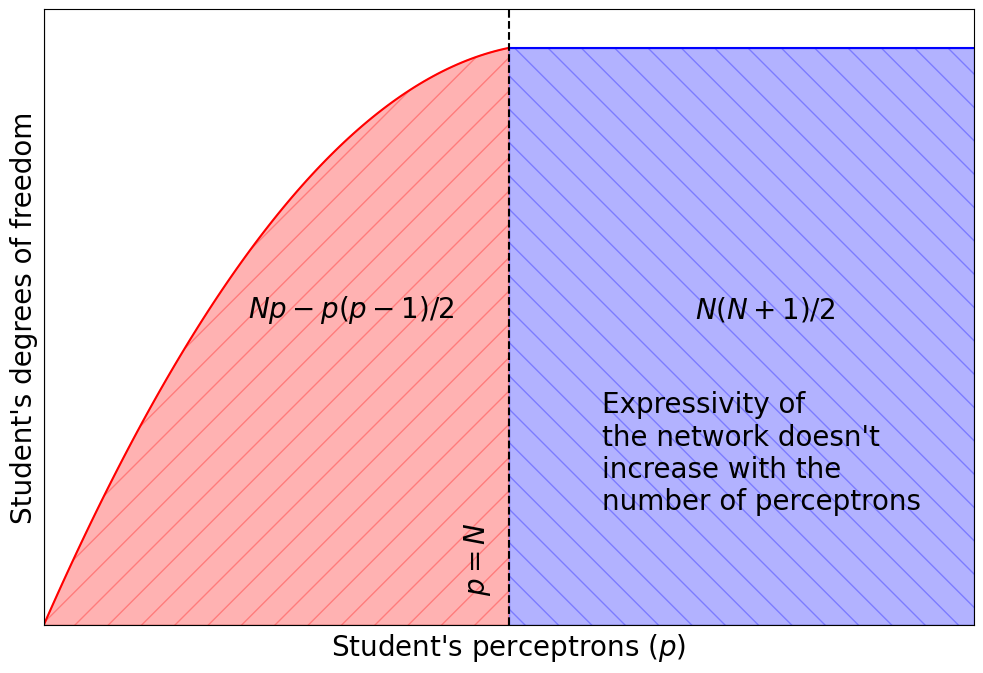}
    \caption{Degrees of freedom of the student network $DOF$ as a function of the hidden layer width $p$.}
    \label{fig:student_dof}
\end{figure}

The effective number of degrees of freedom is then given by
\begin{equation*}
    \textit{Student's DOF} = Np - p(p-1)/2 + (p-N)(p-N-1)/2 = N(N+1)/2.
\end{equation*}
This corresponds to the total degrees of freedom of all the symmetric matrices in $\mathbb{R}^{N \times N}$ and therefore implies that, \textit{for $p > N$, increasing the hidden layer width does not increase the network expressivity}. A summary graph is show in Figure \ref{fig:student_dof}.

\subsubsection{Dimension of the solutions manifold.} By using similar arguments of degrees of freedom of the networks, we want to re-obtain the results on the dimension of the solution manifold. In order to have a zero error network $\bar{W}$, we must have
\begin{equation*}
G_{\bar{W}} = \bar{W}^{T} \bar{W} = \frac{p}{p^{*}} T^{T} T = \frac{p}{p^{*}} G_{T}
\end{equation*}

which, in particular, implies that $\text{rank}(W) = p^{*}$. Given that a solution $\bar{W}$ exists ($p \geq \min(p^{*}, N)$), all the other solutions can be obtained by applying a rotation to this solution. Here we want to quantify the dimension of this manifold. We notice that this quantity is different from the one evaluated above: here we are fixing the matrix $G_{\bar{W}} = \frac{p}{p^{*}} G_{\bm{W}^{*:}}$ and looking at the dimension of the subspace of the $\bar{W}$ matrices that correspond to the same $G_{\bar{W}}$. Because of this, the solution manifold will be at most equal to the dimension of $\bar{W}$, i.e., $\textit{Manifold dimension} \in [0, p N]$. The dimension of the manifold can be written as a function of the network parameters by considering that there are:

\begin{itemize}
    \item $p(p-1)/2$ rotations that leave the network unchanged
    \item Only $\min(p^{*}, N)$ of the columns of $W$ are independent $\Rightarrow  (p-\min(p^{*}, N))(p-\min(p^{*}, N)-1)/2$ rotations act on the subspace formed by the remaining $(p-\min(p^{*}, N))$ vectors, leaving $W$ itself unchanged.
\end{itemize}

\begin{figure}[h!]
    \centering
    \includegraphics[width=0.5\linewidth]{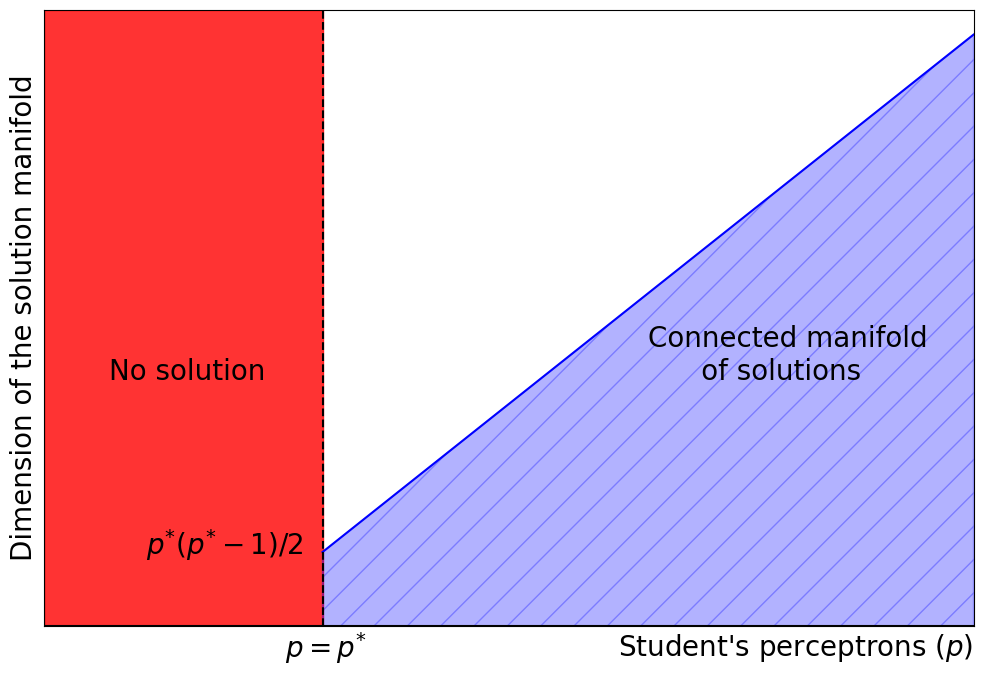}
    \caption{Dimension of the solution manifold as a function of the student network number of hidden layer neurons.}
    \label{fig:manifold_dimension}
\end{figure}

So the dimension of the solution manifold is given by:
\begin{equation*}
    \textit{Manifold dimension} = p(p-1)/2 - (p-\min(p^{*}, N))(p-\min(p^{*}, N) - 1)/2,
\end{equation*}
which for $p \leq N$ is equal to:
\begin{equation*}
    \textit{Manifold dimension} = p(p-1)/2 - (p-p^{*})(p -p^{*} - 1)/2 = pp^{*} - \frac{p^{*}(p^{*}+1)}{2}.
\end{equation*}
The results are summarized in Figure \ref{fig:manifold_dimension}. We can consider the classical phase retrieval problem $p^{*}$ as a consistency check. In this case, the manifold dimension, for $p \geq p^{*}$, is simply given by $p - 1$. Indeed, in this case the solutions include all the configurations where the students are aligned to the teacher vector, but the norms of $(p - 1)$ students are not fixed (the last one has to be chosen in such a way that the zero error condition is met).

\end{document}